\documentclass[lettersize,journal]{IEEEtran}
\usepackage{amsmath,amsfonts}
\usepackage{algorithmic}
\usepackage{algorithm}
\usepackage{array}
\usepackage[caption=false,font=normalsize,labelfont=sf,textfont=sf]{subfig}
\usepackage{textcomp}
\usepackage{stfloats}
\usepackage{url}
\usepackage{verbatim}
\usepackage{graphicx}
\usepackage{cite}
\begin{document}
\title{Joint Design of ISAC Waveform under PAPR Constraints }

{\author{Yating Chen, Cai Wen*, Yan Huang*, Le Liang, Jie Li, Hui Zhang, Wei Hong

        % <-this % stops a space
\thanks{*The corresponding author, email: wencai33@163.com, yellowstone0636@hotmail.com

Yating Chen, Yan Huang, HuiZhang and Wei Hong are with the State Key Laboratory of Millimeter Waves, School of Information Science and Engineering, Southeast University, Nanjing, China.

Cai Wen is with the School of Information Science and Technology, Northwest University, Xi'an, China.

Le Liang is with the National Mobile Communications Research Laboratory, Nanjing  211100, China.

Jie Li is with the College of Electronic and Information Engineering,
Nanjing University of Aeronautics and Astronautics, Nanjing 211100, China.

}}}% <-this % stops a space

\maketitle

\begin{abstract}

In this paper, we formulate the precoding problem of integrated sensing and communication (ISAC) waveform as a non-convex quadratically constrainted quadratic program (QCQP), in which the weighted sum of communication  multi-user interference (MUI) and the gap between dual-use waveform and ideal radar waveform is minimized with peak-to-average power ratio (PAPR) constraints. We propose an efficient algorithm based on alternating direction method of multipliers (ADMM), which is able to decouple multiple variables and provide a closed-form solution for each subproblem. In addition, to improve the sensing performance in both spatial and temporal domains, we propose a new criteria to design the ideal radar waveform, in which the beam pattern is made similar to the ideal one and the integrated sidelobe level of the ambiguity function in each target direction is minimized in the region of interest. The limited memory Broyden-Fletcher-Goldfarb-Shanno (L-BFGS) algorithm is applied to the design of the ideal radar waveform which works as a reference in the design of the dual-function waveform. Numerical results indicate that the designed dual-function waveform is capable of offering good communication quality of service (QoS) and sensing performance.

\begin{IEEEkeywords}
Integrated sensing and communication, Waveform design, MIMO, OFDM, PAPR, Ambiguity function.
\end{IEEEkeywords}

\end{abstract}

\section{introduction}
\label{s1}

Integrated sensing and communication (ISAC) has become a research hotspot for recent years. It aims to merge radar and communication into a single system. The application of ISAC is expected to considerably improve the whole system in terms of equipment size, weight, cost, power consumption, system performance, and spectrum efficiency. In addition, ISAC alleviates electromagnetic compatibility problem in the increasingly complicated electromagnetic environment, where a surging amount of electronic equipment operates in the same platform \cite{1}. Moreover, the high transmission power and strong directivity of radar can enhance the quality, security, and capacity of the communication system. The communication system is also able to assist with low-latency information transmission between radars. It is envisioned that ISAC will be extensively used in many scenarios, like vehicle-to-everything (V2X), smart home, industrial upgrading, and human-computer interaction \cite{2}, where V2X is our focus in this paper. With the help of V2X, vehicles are capable of choosing routes, complying with traffic regulations and avoiding hazards without manual intervention \cite{3}. Radar and communication are both indispensable components for V2X, enabling vehicles to sense the surroundings and exchange information with neighboring vehicles or control nodes.

Generally, the preliminary waveform design approaches of ISAC can be classified into three categories, namely, communication-centric design, radar-centric design, and joint design. Communication-centric approaches use general communication signal for sensing \cite{4,5,6,7,8}. This type of approaches guarantees the performance of communication at the expense of certain sensing performance. Radar-centric methods embed communication data in the radar waveform \cite{9,10,11,12,13,14}. These methods usually bring unsatisfactory quality of communication due to limited data rates without much degradation of sensing performance. Joint design schemes strike a great balance between both functions \cite{15,16,17,18,19}. Although the complexity of joint design is generally slight higher, the joint design methods can achieve scalable performance trade-off between the two functionalities and adjust flexibly the radar and communication performance according to the practical situation. Higher degree of freedom is available in the joint design, because it is not limited by existing radar waveforms or communication waveforms. To obtain satisfactory performance both in radar and communication, we focus on joint design in this paper.
 
Many kinds of radar or communication waveforms can be applied in joint design of the ISAC waveform, among which we choose orthogonal frequency division duplexing (OFDM) waveform. OFDM is commonly used in the field of communication due to its advantages of high spectral efficiency, robustness against multi-path fading, flexible multi-carrier modulation, and ease for synchronization\cite{20}. In the radar field, OFDM can also be utilized for sensing \cite{21}. Its application in both fields lays the foundation for introducing it into the field of ISAC. However, in the integrated sensing and communication system with a single antenna, OFDM waveforms have downsides. Different from the continous OFDM waveform in communication, pulse OFDM waveform is used in ISAC for fusion with radar \cite{22}. With pulse OFDM waveform, the data transmission rate is limited by the frequency of the radar transmitted pulse. Although the data transmission rate can be improved by increasing the pulse repetition rate, this practice will reduce the maximum unambiguous range. To fix this problem, multiple-input multiple-output (MIMO) is incorporated into our design of the ISAC waveform. MIMO can increase the data rate, thus promoting the integration of radar and communication into practical application. 

In recent years, many researchers have proposed joint design methods based on OFDM. In \cite{23}, it presents a subspace-based method for OFDM-based system to jointly estimate range and Doppler shift . Lorenzo Gaudio et al. study the performance of an ISAC system by using two digital modulation formats, namely, the OFDM and the orthogonal time frequency space (OTFS) in terms of mean square error (MSE) and achievable rate in \cite{24}. In \cite{25}, a phase-coded OFDM (PC-OFDM) integrated waveform based on cyclic shifts of m-sequence is proposed, where corresponding time shift is controlled by communication data. In \cite{26}, it proposes three different power minimization-based waveform design criteria to minimize the worst-case radar transmitted power, which are constrained by a mutual information requirement for target characterization and a minimum capacity threshold for communication system. Yongjun Liu et al. propose an adaptive weighted-optimal and Pareto-optimal design approach, considering Cramér-Rao bounds (CRBs) and communication channel capacity in \cite{27}. A multi-objective optimization is built to design ISAC waveforms with radar sensing performance, communication channel capacity and peak-to-average power ratio (PAPR) in \cite{28}. Although it restrains peak sidelobe level (PSL) and PAPR in the design, it does not give a specific definition of them. Some work further employs MIMO-OFDM waveforms. Work in \cite{29} achieves tradeoff between radar and communication by changing the number of subcarriers marked as shared or private. A weighted objective function on both communication and radar performance metrics under power and PAPR constraints is investigated in \cite{30}. While giving specific definition of the PAPR, it formulates one PAPR for all antennas which is not practical. The restriction of the PAPR is brought by the power amplifier, and thus should be defined respectively for every transmission antenna because each antenna has its own power amplifier. Among all the previous proposals, the PAPR and ambiguity function are not given enough attention in the design of ISAC waveform.

Since the dynamic range of the general power amplifier is limited, MIMO-OFDM signal with a high PAPR easily falls into the nonlinear region of power amplifiers, resulting in nonlinear distortion. Restraining PAPR under a certain level in the optimization avoids signal distortion, even with low-cost power amplifiers. In the V2X scenario, the ISAC systems tend to be densely deployed in urban environments with strong clutters. For this reason, the ambiguity function should be intentionally controlled to avoid performance deterioration of sensing caused by interference from nearby vehicles \cite{31}. 

By considering the above problems, we make some improvements in this paper. We first restrict the ambiguity function in the region of interest, which is embedded in the design of the ideal radar waveform. Beam pattern is also considered in the design of the ideal radar waveform. Herein, the limited memory Broyden-Fletcher-Goldfarb-Shanno (L-BFGS) algorithm is applied to the design of the ideal radar waveform which works as a reference in the design of ISAC waveform. With the ideal radar waveform calculated in advance, the optimization problem minimizes multi-user interference (MUI), as well as the gap between the ideal radar waveform and the ISAC waveform. We define a PAPR for each antenna and set PAPRs as constraints along with the total power constraint in optimization. The whole optimization is solved using the alternating direction method of multipliers (ADMM) for fast convergence. The ADMM decouples multiple variables and the subproblem of each variable has a closed-form solution, thus achieving high processing speed.

The contributions of this paper are highlighted as follows. Firstly, we build the model under the scenario of V2X. Considering the complicated electromagnetic environment, the integrated sidelobe level (ISL) of ambiguity function is restricted. Secondly, the PAPR corresponding to each antenna is set as a constraint in the optimization. Thirdly, to make the design more practical, we take into account oversampling, in addition to the scenario with Nyquist sampling.

The rest contents of this paper are organized as follows. Section II describes the system model and signal model, including the ISAC waveform, the performance metric of communication, the performance metric of radar, the definition of PAPRs, and the problem formulation. The design of the ideal radar waveform with L-BFGS as well as the design of ISAC waveform with ADMM are introduced detailedly in section III. In Section IV, numourous simulation results are provided. In the end, we give a conclusion in Section VI.

{\textbf{Notations:}} 

We use $\Bbb{R}$ and $\Bbb{C}$ to represent the set of real numbers and complex numbers respectively. Vectors and matrices are denoted by boldface lowercase and boldface uppercase letters, and scalars are denoted by normal font. The superscripts ${{\left(  \cdot   \right)}^{\text{T}}}$ and ${{\left( \cdot  \right)}^{\text{H}}}$ denote the transpose and conjugate transpose respectively. ${{\left\| \cdot  \right\|}_{2}}$  represents  ${{l}_{2}}$-norm. $\text{real}\left( \cdot  \right)$ and $\left| \cdot  \right|$ indicates the real part and amplitude of a complex number. $\text{vec}\left( \cdot  \right)$ represents vectorization operation of a matrix. $\text{tr}\left( \cdot  \right)$ gives the trace of a matrix. $\otimes $ denotes the Kronecker product. ${{\lambda }_{\min }}\left( \cdot  \right)$ represents the minimum eigenvalue of a matrix. ${{\mathbf{I}}_{\text{n}}}$ is an $n\times n$ identity matrix. And $j =\sqrt{-1}$.

\section{System and Signal Model}
\label{s2}

%\subsection{System Model}
%\label{s2-1}

Assume that the ISAC system is installed on a vehicle or a base station along the road, the users and targets we focus on are vehicles on road. Suppose that the system transmits communication data to $U$ users, i.e., vehicle targets for the system to detect, and there are ${{N}_{t}}$ antennas set in the system for transmitting the ISAC waveform. The ISAC system adopts the basic signal form of MIMO-OFDM, which has ${{N}_{s}}$ subcarriers in the frequency domain and accordingly ${{N}_{s}}$ OFDM symbols in the time domain. We utilize the cyclic prefix (CP) to eliminate inter-carrier interference (ICI) and inter-symbol interference (ISI) which is common in systems adopting OFDM \cite{32} and the length of CP is denoted as ${{N}_{\text{cp}}}$.

%\subsection{signal model}
%\label{s2-2} 

\subsection{Signal model with Nyquist sampling}
\label{s2-1}
The original transmission symbol vector is denoted as ${{\mathbf{s}}_{\text{D}}}\in {{\mathbb{C}}^{{{N}_{\text{s}}}U\times 1}}$ given by
\begin{equation}
{{\mathbf{s}}_{\text{D}}}={{\left[ {{\mathbf{s}}_{\text{D1}}}^{\text{T}},{{{\mathbf{s}}_{\text{D2}}}}^{\text{T}},\cdots ,\mathbf{s}_{\text{D}{{\text{N}}_{\text{s}}}}^{\text{T}} \right]}^{\text{T}}},
\end{equation}
where ${{\mathbf{s}}_{\text{Dn}}}\in {{\mathbb{C}}^{U\times 1}}$ represents the symbol vector carried by the \textit{n}-th subcarrier and its element ${{\left[ {{\mathbf{s}}_{\text{Dn}}} \right]}_{\text{i}}}$ corresponds to the symbol transmitted to the \textit{i}-th user carried by the \textit{n}-th subcarrier. In the frequency domain, the symbol vector is precoded by the matrix $\mathbf{P}\in {{\mathbb{C}}^{{{N}_{s}}{{N}_{t}}\times {{N}_{s}}U}}$ written as 
\begin{equation}
\mathbf{P}=\text{diag}({{\mathbf{P}}_{1}},{{\mathbf{P}}_{2}},\cdots ,{{\mathbf{P}}_{{{\text{N}}_{\text{s}}}}}),
\end{equation}
where ${{\mathbf{P}}_{\text{n}}}\in {{\mathbb{C}}^{{{N}_{t}}\times U}}$ is the precoding matrix designed for the \textit{n}-th subcarrier. The \textit{u}-th column of ${{\mathbf{P}}_{\text{n}}}$ precodes the data carried by the \textit{n}-th subcarrier, which is transmitted to the \textit{u}-th user. The precoded symbol vector is given by $\mathbf{x}$, that is,
\begin{equation}
\mathbf{x}=\mathbf{P}{{\mathbf{s}}_{\text{D}}},
\end{equation}
where
\begin{equation}
\mathbf{x}={{\left[ {{\mathbf{x}}_{1}}^{\text{T}},{{\mathbf{x}}_{2}}^{\text{T}},\cdots ,{{\mathbf{x}}_{{{\text{N}}_{\text{s}}}}}^{\text{T}} \right]}^{\text{T}}},
\end{equation}
and ${{\mathbf{x}}_{\text{n}}}\in {{\mathbb{C}}^{{{N}_{\text{t}}}\times 1}}$ is the precoded data vector carried by the \textit{n}-th subcarrier as ${{\mathbf{x}}_{\text{n}}}={{\mathbf{P}}_{\text{n}}}{{\mathbf{s}}_{\text{Dn}}}$.

The precoded symbol vector should be transformed to time domain with inverse discrete Fourier transform (IDFT). Define the normalized DFT matrix as $\mathbf{F}\in {{\mathbb{C}}^{{{N}_{s}}\times {{N}_{s}}}}$ and its \textit{(i,j)}-th element is
\begin{equation}
{{\mathbf{F}}(i,j)}=\frac{1}{\sqrt{{{N}_{s}}}}{{e}^{-j\frac{2\pi }{{{N}_{s}}}(i-1)(j-1)}}.
\end{equation}
%the precoded symbol  We further use the following matrix multiplication to denote the processing of IDFT, which can be written as
%\begin{equation}
%	{{\mathbf{F}}^{H}}\otimes {{\mathbf{I}}_{{{N}_{t}}}}%\in {{\mathbb{C}}^{{{N}_{s}}{{N}_{t}}\times {{N}_{s}}{{N}_{t}}}}.
%\end{equation}

Then the transmission signal vector in time domain is $\mathbf{s}$ given by
\begin{equation}
	\mathbf{s}=({{\mathbf{F}}^{\text{H}}}\otimes {{\mathbf{I}}_{{{\text{N}}_{\text{t}}}}})\mathbf{x}.
\end{equation}

The vector can be divided into ${{N}_{\text{s}}}$ blocks with each block containing ${{N}_{\text{t}}}$ rows as follows
\begin{equation}
	\mathbf{s}={{\left[\mathbf{s}_{1}^{\text{T}},\cdots ,\mathbf{s}_{\rm{{N}_{s}}}^{\text{T}}\right]}^{\text{T}}}.
\end{equation}

 To suppress the influence of ICI and ISI, CP with length of ${{N}_{\text{cp}}}$ is added to the symbol vector. The CP symbol vector in time domain is denoted as ${{\text{s}}_{\text{cp}}}\in {{\mathbb{C}}^{{{N}_{\text{cp}}}{{N}_{\text{t}}}\times 1}}$ which replicates the last ${{N}_{\text{cp}}}$ symbols of the effective symbols. It can be written as
\begin{equation}
{{\mathbf{s}}_{\text{cp}}}={{[{{\mathbf{s}}_{{{\text{N}}_{\text{s}}}\text{-}{{\text{N}}_{\text{cp}}}\text{+1}}},{{\mathbf{s}}_{{{\text{N}}_{\text{s}}}\text{-}{{\text{N}}_{\text{cp}}}\text{+2}}},\cdots ,{{\mathbf{s}}_{{{\text{N}}_{\text{s}}}}}]}^{\text{T}}}.
\end{equation}

We define a selection matrix ${{\mathbf{\Gamma }}_{\text{cp}}}\in {{\mathbb{C}}^{{{N}_{cp}}{{N}_{t}}\times {{N}_{s}}{{N}_{t}}}}$ with its elements being 
\begin{equation}
{{\mathbf{\Gamma }}_{\text{cp}}}(i,j)=\left\{ \begin{matrix}
   1 & j=i+({{N}_{s}}-{{N}_{cp}}){{N}_{t}}  \\
   0 & \text{otherwise}  \\
\end{matrix} \right..
\end{equation}

The operation of selecting CP symbols from effective symbols can be realized as
\begin{equation}
{{\mathbf{s}}_{\text{cp}}}={{\mathbf{\Gamma }}_{\text{cp}}}\mathbf{s}.
\end{equation}

Considering the effective symbol vector and CP, the complete symbol vector is $\mathbf{\dot{s}}$, which is
\begin{equation}
	\mathbf{\dot{s}}={{[{{\mathbf{s}}_{\text{cp}}}^{\text{T}},{{\mathbf{s}}^{\text{T}}}]}^{\text{T}}}.
\end{equation}

For the sake of representation, we define matrix $\mathbf{\Gamma }$ as
\begin{equation}
\mathbf{\Gamma }={{[\mathbf{\Gamma }_{\text{cp}}^{\text{T}},\mathbf{I}_{{{\text{N}}_{\text{s}}}{{\text{N}}_{\text{t}}}}^{\text{T}}]}^{\text{T}}}.
\end{equation}

Therefore, the complete symbol vector can also be rewritten as follows 
\begin{equation}
	\mathbf{\dot{s}}=\Gamma{{\mathbf{s}}}.
\end{equation}

\subsection{Signal model with oversampling}
\label{s2-2}

Oversampling is commonly used in OFDM systems. In practical OFDM systems, it is usually necessary to use oversampling for digital predistortion to avoid serious distortion of the time-domain signal. Meanwhile, PAPR can be measured more accurately in the case of oversampling rather than Nyquist sampling because PAPR is originally defined in continuous time signal. Therefore, we  also consider the case of oversampling in this paper. Oversampling is implemented on the precoded symbol vector, so the original transmission symbol vector ${{\mathbf{s}}_{\text{D}}}$, the precoding matrix $\mathbf{P}$, and the precoded symbol vector $\mathbf{x}$ are the same as before. Assuming that oversampling rate is $\gamma $, we interpolate precoded symbol vector $\mathbf{x}$ with zero to get vector ${{\mathbf{x}}_{\text{os}}}\in {{\mathbb{C}}^{\gamma {{N}_{s}}{{N}_{t}}\times 1}}$ . The vector can be written as

\begin{equation}
\begin{aligned}
  & {{\mathbf{x}}_{\text{os}}}=\left[ \mathbf{x}_{1}^\text{T},\cdots ,\mathbf{x}_{{{N}_{s}}/2}^\text{T},\underbrace{\text{0},\cdots ,\text{0},\cdots ,\text{0}}_{\left( \gamma -1 \right){{N}_{s}}{{N}_{t}}}, \right. \\ 
 & {{\left. \mathbf{x}_{1+{{N}_{s}}/2}^\text{T},\cdots ,\mathbf{x}_{{{N}_{s}}}^\text{T} \right]}^\text{T}} \\ 
\end{aligned}
\end{equation}

Under the situation of oversampling, we define a normalized DFT matrix ${{\mathbf{F}}_{\text{os}}}\in {{\mathbb{C}}^{\gamma {{N}_{s}}\times \gamma {{N}_{s}}}}$  given by
\begin{equation}
{{\mathbf{F}}_{\text{os}}}(i,j)=\frac{1}{\sqrt{\gamma {{N}_{s}}}}{{e}^{-j\frac{2\pi }{\gamma {{N}_{s}}}(i-1)(j-1)}}.
\end{equation}

%The operation of IDFT can be realized by:
%\begin{equation}
%	\mathbf{F}_{os}^{H}\otimes {{\mathbf{I}}_{{{N}_{t}}}}\in {{\mathbb{C}}^{\gamma {{N}_{s}}{{N}_{t}}\times \gamma {{N}_{s}}{{N}_{t}}}}.
%\end{equation}

With the definition above, the transmission symbol vector in time domain is ${{\mathbf{s}}_{\text{os}}}$, which is formulated as
\begin{equation}
{{\mathbf{s}}_{\text{os}}}=(\mathbf{F}_{\text{os}}^{\text{H}}\otimes {{\mathbf{I}}_{{{\text{N}}_{\text{t}}}}}){{\mathbf{x}}_{\text{os}}}.
\end{equation}

The transmission symbol vector ${{\mathbf{s}}_{\text{os}}}$ can be divided into $\gamma {{N}_{s}}$ blocks with each block as %${{\mathbf{\tilde{S}}}_{n}}\in {{\mathbb{C}}^{{{N}_{t}}\times 1}}$:
\begin{equation}
{{\mathbf{s}}_{\text{os}}}={{\left[\mathbf{\tilde{s}}_{1}^{\text{T}},\cdots ,\mathbf{\tilde{s}}_{\text{ }\!\!\gamma\!\!\text{ }{{\text{N}}_{\text{s}}}}^{\text{T}}\right]}^{\text{T}}}.
\end{equation}

The CP symbol vector ${{\mathbf{\tilde{s}}}_{\text{cp}}}\in {{\mathbb{C}}^{\gamma {{N}_{cp}}{{N}_{t}}\times 1}}$ is the same as the last part of ${{\mathbf{s}}_{\text{os}}}$ with CP length of ${\gamma}{{N}_{\text{cp}}}$, expressed as
\begin{equation}
{{\mathbf{\tilde{s}}}_{\text{cp}}}={{[{{\mathbf{\tilde{s}}}_{\text{ }\!\!\gamma\!\!\text{ }{{\text{N}}_{\text{s}}}\text{- }\!\!\gamma\!\!\text{ }{{\text{N}}_{\text{cp}}}\text{+1}}},{{\mathbf{\tilde{s}}}_{\text{ }\!\!\gamma\!\!\text{ }{{\text{N}}_{\text{s}}}\text{- }\!\!\gamma\!\!\text{ }{{\text{N}}_{\text{cp}}}\text{+2}}},\cdots ,{{\mathbf{\tilde{s}}}_{\text{ }\!\!\gamma\!\!\text{ }{{\text{N}}_{\text{s}}}}}]}^{\text{T}}}.
\end{equation}

We define a matrix ${{\mathbf{\tilde{\Gamma }}}_{\text{cp}}}\in {{\mathbb{C}}^{\gamma {{N}_{cp}}{{N}_{t}}\times \gamma {{N}_{s}}{{N}_{t}}}}$ to obtain the CP part from the effective symbol vector ${{\mathbf{s}}_{\text{os}}}$, where
\begin{equation}
{{\mathbf{\tilde{\Gamma }}}_{\text{cp}}}(i,j)=\left\{ \begin{matrix}
   1 & j=i+(\gamma {{N}_{s}}-\gamma {{N}_{cp}}){{N}_{t}}  \\
   0 & \text{otherwise}  \\
\end{matrix} \right..
\end{equation}

Therefore, the CP symbol vector can be expressed as
\begin{equation}
	{{\mathbf{\tilde{s}}}_{\text{cp}}}={{\mathbf{\tilde{\Gamma }}}_{\text{cp}}}{{\mathbf{s}}_{\text{os}}}.
\end{equation}

By combining effective symbols and CP symbols, the complete symbol vector with oversampling is %$\mathbf{\tilde{S}}\in {{\mathbb{C}}^{({{N}_{cp}}+\gamma {{N}_{s}}){{N}_{t}}\times 1}}$  written as
\begin{equation}
\mathbf{\tilde{s}}={{[\mathbf{\tilde{s}}_{\text{cp}}^{\text{T}},\mathbf{s}_{\text{os}}^{\text{T}}]}^{\text{T}}}.
\end{equation}

For simplicity of subsequent equations, we define $\mathbf{\tilde{\Gamma }}$ as
\begin{equation}
\mathbf{\tilde{\Gamma }}={{[\mathbf{\tilde{\Gamma }}_{\text{cp}}^{\text{T}},\mathbf{I}_{\text{ }\!\!\gamma\!\!\text{ }{{\text{N}}_{\text{s}}}{{\text{N}}_{\text{t}}}}^{\text{T}}]}^{\text{T}}}.
\end{equation}

By using $\mathbf{\tilde{\Gamma }}$ and ${{\mathbf{s}}_\text{os}}$, the complete symbol vector with oversampling can be expressed as
\begin{equation}
\mathbf{\tilde{s}}=\tilde{\Gamma }{{\mathbf{s}}_{\text{os}}}.
\end{equation}

\subsection{Performance metric of communication}
\label{s2-3}

In millimeter wave band, we adopt Rician channel model. A Rician channel obeys the probability density function (PDF) of \cite{33}
\begin{equation}
    p(x)=\frac{x }{\sigma _\text{n}^{2}}{{e}^{\frac{{{x}^{2}}+{{A}^{2}}}{2\sigma _\text{n}^{2}}}}{{I}_{0}}(\frac{Ax}{\sigma _\text{n}^{2}}),
\end{equation}
where $A$ is the amplitude of the line-of-sight (LOS) path, $\sigma _\text{n}$ is the standard deviation of the scattered multipath amplitudes and ${I}_{0}(y)$ is the 1st kind modified Bessel function of zero order \cite{34}. The Rician factor $K$ is defined as
\begin{equation}
K=\frac{{{A}^{2}}}{2\sigma _\text{n}^{2}}
\end{equation}
The direction of the dominant signal is the same as radar channel. We denote channel matrix $\mathbf{H}\in {{\mathbb{C}}^{{{N}_{s}}U\times {{N}_{s}}{{N}_{t}}}}$ as 
\begin{equation}
\mathbf{H}=\text{diag}({{\mathbf{H}}_{1}},\cdots ,{{\mathbf{H}}_{{{\text{N}}_{\text{s}}}}}),
\end{equation}
where ${{\mathbf{H}}_{\text{n}}}\in {{\mathbb{C}}^{U\times {{N}_{t}}}}$ represents the channel state matrix in frequency domain corresponding to the \textit{n}-th subcarrier. It is defined specifically as
\begin{equation}
{{\mathbf{H}}_{\text{n}}}=\sum\limits_{t=0}^{T-1}{{{{\mathbf{\tilde{H}}}}_{\text{t}}}{{e}^{-j\frac{2\pi t(n-1)}{{{N}_{s}}}}}},
\end{equation}
where $T-1$ is the memory of the fading channel. ${{\mathbf{\tilde{H}}}_{\text{t}}}\in {{\mathbb{C}}^{U\times {{N}_{t}}}}$ is the channel state matrix for the \textit{t}-th tap obeying Rician distribution.
 
The symbol vector received by communication receiver is %$\mathbf{Y}\in {{\mathbb{C}}^{{{N}_{s}}U\times 1}}$ with 
\begin{equation}
	\mathbf{y}=\mathbf{H}\mathbf{x}+\mathbf{z},
\end{equation}
where $\mathbf{z}$ represents additive Gaussian white noise and each element of $\mathbf{z}$ independently and identically obeys Gaussian distribution.

MUI is a significant metric of communication in a MIMO-OFDM system. It is proved that MUI reduces the achievable sum rate and increases the average symbol error rate (SER) thus deteriorating communication performance \cite{35}. Based on the definition above, MUI is represented as $\mathbf{H}\mathbf{x}-{{\mathbf{s}}_{\text{D}}}$. From the perspective of communication, MUI should be minimized in the optimization to improve the performance of the proposed ISAC waveform. 

\subsection{Performance metric of radar}
\label{s2-4}
\subsubsection{Beam pattern}
\label{s2-4-1}

Furthermore, by considering the complicated electromagnetic environment around the target to be detected, the beam pattern is crucial for sensing accuracy. The beam pattern can be used to describe the synthesis of beam in space. The direction of beam, the position and amplitude of sidelobes in space can be observed through beam pattern. Hence, the beam pattern is intentionally controlled in the design to ensure satisfactory sensing performance. The spatial covariance matrix of the transmitted signal is ${{\mathbf{R}}_{\text{n}}}\in {{\mathbb{C}}^{{{N}_{t}}\times {{N}_{t}}}}$ which corresponds to the $n$-th OFDM symbol. The beam pattern is defined as a function of detection angle  $\theta $ written as
\begin{equation}
b(\theta )=\frac{1}{{{N}_{s}}}\sum\limits_{n=1}^{{{N}_{s}}}{{{\mathbf{a}}^{\text{H}}}(\theta )}{{\mathbf{R}}_{\text{n}}}\mathbf{a}(\theta ).
\end{equation}

In the definition, $\mathbf{a}(\theta )\in {{\mathbb{C}}^{{{N}_{t}}\times 1}}$ is the transmit steering vector with definition of 
\begin{equation}
\begin{aligned}
	\mathbf{a}(\theta )=&\left[ {{e}^{j(1-\frac{{{N}_{t}}}{2})\pi \sin \theta }},{{e}^{j(2-\frac{{{N}_{t}}}{2})\pi \sin \theta }}, \right. \\ 
	& {{\left. \cdots ,{{e}^{j({{N}_{t}}-\frac{{{N}_{t}}}{2})\pi \sin \theta }} \right]}^{\text{T}}}.  
\end{aligned}
\end{equation}

\subsubsection{Ambiguity function}
\label{s2-4-2}

Ambiguity function is a significant metric to evaluate the Doppler and range resolutions \cite{36}. It can also be employed to measure the interference mitigation capability of a radar system \cite{31}. It is commonly applied for performance analysis of radar systems, measuring the ability of radar to distinguish targets with different distances and speeds. The ambiguity function, $\chi (\theta ,k,f)$, of a MIMO radar is a 3-D function of $\theta $, $k$ and $f$ \cite{37}, where $\theta $ is the detection angle, $k$ is range bin and $f$ is normalized Doppler frequency.

The signal transmitted towards the direction $\theta $ is denoted by ${{\mathbf{s}}_{\text{v}}}$, which can be expressed as
\begin{equation}
{{\mathbf{s}}_{\text{v}}}(\theta )={{[{{\mathbf{\dot{s}}}^{\text{T}}}({{\mathbf{I}}_{{{\text{N}}_{\text{s}}}\text{+}{{\text{N}}_{\text{cp}}}}}\otimes \mathbf{a})]}^{\text{T}}}.
\end{equation}

The matrix corresponding to the delay $k$ is ${{\mathbf{J}}_{\text{k}}}\in {{\mathbb{C}}^{({{N}_{s}}+{{N}_{cp}})\times ({{N}_{s}}+{{N}_{cp}})}}$, written as
\begin{equation}
{{\mathbf{J}}_{\text{k}}}(i,j)=\left\{ \begin{matrix}
   1 & i-j=k  \\
   0 & i-j\ne k  \\
\end{matrix} \right..
\end{equation} 

The matrix corresponding to the normalized Doppler frequency $f$ is ${{\mathbf{D}}_{\text{f}}}\in {{\mathbb{C}}^{({{N}_{s}}+{{N}_{cp}})\times ({{N}_{s}}+{{N}_{cp}})}}$, defined as
\begin{equation}
{{\mathbf{D}}_{\text{f}}}=\text{diag}({{e}^{j2\pi f}},\cdots ,{{e}^{j2\pi ({{N}_{s}}+{{N}_{cp}})f}}).
\end{equation} 

In the definition of the ambiguity function, the matrix ${{\mathbf{J}}_{\text{k}}}$ adds time shift to the signal while the matrix ${{\mathbf{D}}_{\text{f}}}$ adds Doppler frequency shift to the signal, which is consistent with the common definition of ambiguity function. With the definition of ${{\mathbf{J}}_{\text{k}}}$ and ${{\mathbf{D}}_{\text{f}}}$, we can derive the ambiguity function of the detection angle $\theta $ as \cite{38}
\begin{equation}
\chi (\theta ,k,f)={{\left| {{\mathbf{s}}_{\text{v}}}{{(\theta )}^{\text{H}}}{{\mathbf{J}}_{\text{k}}}{{\mathbf{D}}_{\text{f}}}{{\mathbf{s}}_{\text{v}}}(\theta ) \right|}^{2}}.
   \label{ambiguity}
\end{equation}

In the case of oversampling, the transmission symbol vector towards the detection angle of $\theta $ is presented as  ${{\mathbf{\tilde{s}}}_{\text{v}}}(\theta )$, which is presented as 
\begin{equation}
{{\mathbf{\tilde{s}}}_{\text{v}}}(\theta )={{[{{\mathbf{\tilde{s}}}^{\text{T}}}({{\mathbf{I}}_{\text{ }\!\!\gamma\!\!\text{ }{{\text{N}}_{\text{s}}}\text{+}{\text{ }\!\!\gamma\!\!\text{ }{\text{N}}_{\text{cp}}}}}\otimes \mathbf{a}(\theta ))]}^{\text{T}}}.
\end{equation}

Similar to the derivation above, we redefine ${{\mathbf{\tilde{J}}}_{\text{k}}}\in {{\mathbb{C}}^{(\gamma {{N}_{s}}+\gamma {{N}_{cp}})\times (\gamma {{N}_{s}}+\gamma {{N}_{cp}})}}$ and ${{\mathbf{\tilde{D}}}_{\text{f}}}\in {{\mathbb{C}}^{(\gamma {{N}_{s}}+\gamma {{N}_{cp}})\times (\gamma {{N}_{s}}+\gamma {{N}_{cp}})}}$ to adjust to the oversampled symbol vector with
\begin{equation}
{{\mathbf{\tilde{J}}}_{\text{k}}}(i,j)=\left\{ \begin{matrix}
   1 & i-j=k  \\
   0 & i-j\ne k  \\
\end{matrix} \right.
\end{equation}
and
\begin{equation}
{{\mathbf{\tilde{D}}}_{\text{f}}}=\text{diag}({{e}^{j2\pi f}},\cdots ,{{e}^{j2\pi (\gamma {{N}_{s}}+\gamma {{N}_{cp}})f}}).
\end{equation}

It should be noticed that the normalized Doppler frequency $f$ here is different from the frequency under Nyquist sampling rate. The Doppler frequency is normalized by the sampling frequency in the specific situation.

The ambiguity function of detection angle of $\theta $ with oversampling is
\begin{equation}
\chi (\theta ,k,f)={{\left| {{{\mathbf{\tilde{s}}}}_{\text{v}}}{{(\theta )}^{\text{H}}}{{{\mathbf{\tilde{J}}}}_{\text{k}}}{{{\mathbf{\tilde{D}}}}_{\text{f}}}{{{\mathbf{\tilde{s}}}}_{\text{v}}}(\theta ) \right|}^{2}}.
\end{equation}

\subsection{PAPR}
\label{s2-5}
\subsubsection{PAPR with Nyquist-sampling}
\label{s2-5-1}

We define an independent PAPR for every transmission antenna because each antenna has its own power amplifier in practical applications. Since there are ${{N}_{\text{t}}}$ antennas, we define ${{N}_{\text{t}}}$ PAPRs with ${\rm{PAPR}_{l}}$ relative to the \textit{l}-th antenna.

For subsequent calculation, we define selection matrix ${{\mathbf{C}}_{\text{l}}}\in {{\mathbb{C}}^{{{N}_{s}}\times {{N}_{s}}{{N}_{t}}}}$ as
\begin{equation}
{{\mathbf{C}}_{\text{l}}}(i,j)=\left\{ \begin{matrix}
   1 & j=(i-1){{N}_{t}}+l  \\
   0 & \text{otherwise}  \\
\end{matrix} \right..
\end{equation}

Leveraging the selection matrix, the ${{N}_{s}}$ OFDM symbols transmitted by the \textit{l}-th antenna can be extracted from the symbol vector. The symbol vector of the \textit{l}-th antenna is ${{\mathbf{\hat{s}_{\text{l}}}}}$, which can be presented as
\begin{equation}
{{\mathbf{\hat{s}_{\text{l}}}}}={{\mathbf{C}}_{\text{l}}}\mathbf{s}.
\end{equation}

As the CP symbols are the same as the end of the original symbols, CP symbols can be neglected in the calculation of PAPR. According to the definition of PAPRs, the PAPR of the \textit{l}-th antenna is derived as
\begin{equation}
\begin{aligned}
   \text{PAP}{{\text{R}}_{\text{l}}}&=\frac{\underset{n}{\mathop{\max }}\,{{\left| {{[{{{\mathbf{\hat{s}}}}_{\text{l}}}]}_{\text{n}}} \right|}^{2}}}{\frac{1}{{{N}_{s}}}\left\| {{{\mathbf{\hat{s}}}}_{\text{l}}} \right\|_{2}^{2}} \\ 
 & =\frac{\underset{n}{\mathop{\max }}\,{{\left| {{[{{\mathbf{C}}_{\text{l}}}({{\mathbf{F}}^{\text{H}}}\otimes {{\mathbf{I}}_{{{\text{N}}_{\text{t}}}}})\mathbf{x}]}_{\text{n}}} \right|}^{2}}}{\frac{1}{{{N}_{s}}}\left\| [{{\mathbf{C}}_{\text{l}}}({{\mathbf{F}}^{\text{H}}}\otimes {{\mathbf{I}}_{{{\text{N}}_{\text{t}}}}})\mathbf{x}] \right\|_{2}^{2}}.  
\end{aligned}
\end{equation}

\subsubsection{PAPR with oversampling}
\label{s2-4-2}

In a further consideration of oversampling, the derivation of PAPRs is similar to that of Nyquist sampling. To represent the PAPR, we first get the $\gamma {{N}_{s}}$ OFDM symbols of the \textit{l}-th antenna, which is presented as
\begin{equation}
{{\mathbf{\dot{s}}}_{\text{l}}}={{\mathbf{\tilde{C}}}_{\text{l}}}{{\mathbf{s}}_{\text{os}}}.
\end{equation}

The choosing matrix ${{\mathbf{\tilde{C}}}_{\text{l}}}\in {{\mathbb{C}}^{\gamma {{N}_{s}}\times \gamma {{N}_{s}}{{N}_{t}}}}$ in the above equation is defined as 
\begin{equation}
{{\mathbf{\tilde{C}}}_{\text{l}}}(i,j)=\left\{ \begin{matrix}
   1 & j=(i-1){{N}_{t}}+l  \\
   0 & \text{otherwise}  \\
\end{matrix} \right..
\end{equation}

According to the definition of PAPRs, the PAPR of the \textit{l}-th antenna is 
\begin{equation}
\text{PAP}{{\text{R}}_{\text{l}}}=\frac{\underset{n}{\mathop{\max }}\,{{\left| {{[{{{\mathbf{\tilde{C}}}}_{\text{l}}}(\mathbf{F}_{\text{os}}^{\text{H}}\otimes {{\mathbf{I}}_{{{\text{N}}_{\text{t}}}}}){{\mathbf{x}}_{\text{os}}}]}_{\text{n}}} \right|}^{2}}}{\frac{1}{\gamma {{N}_{s}}}\left\| [{{{\mathbf{\tilde{C}}}}_{\text{l}}}(\mathbf{F}_{\text{os}}^{\text{H}}\otimes {{\mathbf{I}}_{{{\text{N}}_{\text{t}}}}}){{\mathbf{x}}_{\text{os}}}] \right\|_{2}^{2}}.
\end{equation}

For the convenience of following optimization, we redefine a DFT matrix ${{\mathbf{\tilde{F}}}_{\text{os}}}\in {{\mathbb{C}}^{{{N}_{s}}\times \gamma {{N}_{s}}}}$ with

\begin{equation}
\begin{aligned}
  & {{{\mathbf{\tilde{F}}}}_{\text{os}}}(i,j)= \\ 
 & \left\{ \begin{matrix}
   \frac{1}{\sqrt{{{N}_{s}}}}{{e}^{-j\frac{2\pi }{\gamma {{N}_{s}}}(i-1)(j-1)}}  \\
   (i=1,\cdots ,\frac{{{N}_{s}}}{2},j=1,\cdots ,\gamma {{N}_{s}})  \\
   \frac{1}{\sqrt{{{N}_{s}}}}{{e}^{-j\frac{2\pi }{\gamma {{N}_{s}}}\left[ (\gamma -1){{N}_{s}}+(i-1) \right](j-1)}}  \\
   (i=\frac{{{N}_{s}}}{2}+1,\cdots ,{{N}_{s}},j=1,\cdots ,\gamma {{N}_{s}})  \\
\end{matrix} \right. .\\ 
\end{aligned}\\ 
\end{equation}

It is proved that the equation below is satisfied \cite{30}:
\begin{equation}
(\mathbf{F}_{\text{os}}^{\text{H}}\otimes {{\mathbf{I}}_{{{\text{N}}_{\text{t}}}}}){{\mathbf{x}}_{\text{os}}}=(\mathbf{\tilde{F}}_{\text{os}}^{\text{H}}\otimes {{\mathbf{I}}_{{{\text{N}}_{\text{t}}}}})\mathbf{x}.
\end{equation}

With the equation, the PAPR can be rewritten as 
\begin{equation}
\text{PAP}{{\text{R}}_{\text{l}}}=\frac{\underset{n}{\mathop{\max }}\,{{\left| {{[{{{\mathbf{\tilde{C}}}}_{\text{l}}}(\mathbf{\tilde{F}}_{\text{os}}^{\text{H}}\otimes {{\mathbf{I}}_{{{\text{N}}_{\text{t}}}}})\mathbf{x}]}_{\text{n}}} \right|}^{2}}}{\frac{1}{\gamma {{N}_{s}}}\left\| [{{{\mathbf{\tilde{C}}}}_{\text{l}}}(\mathbf{\tilde{F}}_{\text{os}}^{\text{H}}\otimes {{\mathbf{I}}_{{{\text{N}}_{\text{t}}}}})\mathbf{x}] \right\|_{2}^{2}}.
\end{equation}

\section{ISAC MIMO-OFDM waveform design}
\label{s3}
\subsection{Design of ideal radar waveform}
\label{s3-1}
The design of ideal radar waveform covers the ambiguity function and beam pattern. On the one hand, we minimize the average value of the ISL over specific Doppler bins and range bins of interest. On the other hand, we minimize the difference between the real beam pattern and the ideal beam pattern. With the two items in consideration, the designed ideal radar waveform guarantees the performance of radar sensing.
\subsubsection{Design of ideal radar waveform with Nyquist sampling}
\label{s3-1-1}

The transmission signal of direction $\theta $ can be rewritten as
\begin{equation}
\begin{aligned}
   {{\mathbf{s}}_{\text{v}}}(\theta )&={{[{{{\mathbf{\dot{s}}}}^{\text{T}}}({{\mathbf{I}}_{{{\text{N}}_{\text{s}}}\text{+}{{\text{N}}_{\text{cp}}}}}\otimes \mathbf{a}(\theta ))]}^{\text{T}}} \\ 
 & ={{({{\mathbf{I}}_{{{\text{N}}_{\text{s}}}\text{+}{{\text{N}}_{\text{cp}}}}}\otimes \mathbf{a}(\theta ))}^{\text{T}}}\mathbf{\dot{s}} \\ 
 & ={{({{\mathbf{I}}_{{{\text{N}}_{\text{s}}}\text{+}{{\text{N}}_{\text{cp}}}}}\otimes \mathbf{a}(\theta ))}^{\text{T}}}\mathbf{\Gamma s} . 
\end{aligned}
\end{equation}
To simplify the expression, we define $\mathbf{G}(\theta )$ as
\begin{equation}
\mathbf{G}(\theta )={{({{\mathbf{I}}_{{{\text{N}}_{\text{s}}}\text{+}{{\text{N}}_{\text{cp}}}}}\otimes \mathbf{a}(\theta ))}^{\text{T}}}\mathbf{\Gamma }.
\end{equation}
Thus, the waveform radiated in direction $\theta$ can be expressed with $\mathbf{G}(\theta )$, which is 
\begin{equation}
{{\mathbf{s}}_{\text{v}}}(\theta )=\mathbf{G}(\theta )\mathbf{s}.
  \label{transmission}
\end{equation}
By plugging \eqref{transmission} into \eqref{ambiguity}, the ambiguity function of detection angle $\theta $ is  ${{\left| {{\mathbf{s}}^{\text{H}}}{{\mathbf{G}}^{\text{H}}}(\theta ){{\mathbf{J}}_{\text{k}}}{{\mathbf{D}}_{\text{f}}}\mathbf{G}(\theta )\mathbf{s} \right|}^{2}}$.

Generally, PSL and ISL are both studied in the research of ambiguity function. ISL is more tractable compared with the PSL metric from an optimization point of view \cite{39}, thus we choose ISL in our optimization. As the total volume of the ambiguity function is fixed, it is unrealizable to achieve a very low ISL over the entire region of delay and Doppler frequency. We only control the ambiguity function of interested time delays and Doppler frequencies, so the design minimizes the summation of these ambiguity function values as
\begin{equation}
	\sum\limits_{\theta \in {{\Omega }_{d}}}{\sum\limits_{(k,f)\in \Delta }{{{\left| {{\mathbf{s}}^{\text{H}}}{{\mathbf{G}}^{\text{H}}}(\theta ){{\mathbf{J}}_{\text{k}}}{{\mathbf{D}}_{\text{f}}}\mathbf{G}(\theta )\mathbf{s} \right|}^{2}}}},
\end{equation}
where ${{\Omega }_{d}}$ represents the collection of the interested detection angles, from which the targets can be detected, and $\Delta $ represents the collection of the interested pairs of time delays and Doppler frequencies.

The design also makes the real beam pattern similar to the ideal one. For each symbol time, there is a beam pattern corresponding to the signal transmitted at the time. In the optimization, we utilize the average beam pattern to approach the ideal one. The similarity between the real beam pattern and the ideal beam pattern is measured by
\begin{equation}
	{{\left[\frac{1}{{{N}_{\text{s}}}+{{N}_{\text{cp}}}}{{\mathbf{s}}^{\text{H}}}{{\mathbf{G}}^{\text{H}}}(\theta )\mathbf{G}(\theta )\mathbf{s}-d(\theta )\right]}^{2}}.
\end{equation}

For each angle, there is a beam pattern value and accordingly a similarity value. In the optimization problem, we sum the values up and minimize the summation.

The design of ideal radar waveform with Nyquist sampling is formulated as
\begin{small}
\begin{equation}
\begin{aligned}
	& \underset{\mathbf{s}}{\mathop{\min }}\,\sum\limits_{\theta \in {{\Omega }_{d}}}{\sum\limits_{(k,f)\in \Delta }{{{\left| {{\mathbf{s}}^{\text{H}}}{{\mathbf{G}}^{\text{H}}}(\theta ){{\mathbf{J}}_{\text{k}}}{{\mathbf{D}}_{\text{f}}}\mathbf{G}(\theta )\mathbf{s} \right|}^{2}}}}+ \\ 
	& {{({{N}_{\text{s}}}+{{N}_{\text{cp}}})}^{2}}\sum\limits_{\theta \in {{\Omega }_{all}}}{{{\left[\frac{{{\mathbf{s}}^{\text{H}}}{{\mathbf{G}}^{\text{H}}}(\theta )\mathbf{G}(\theta )\mathbf{s}}{{{N}_{\text{s}}}+{{N}_{\text{cp}}}}-d(\theta )\right]}^{2}}}, \\ 
\end{aligned}
\end{equation}
\end{small}
where ${{\Omega }_{all}}$ represents the collection of angles covered in the beam pattern. We add a coefficient in front of the second term to make a balance in the optimization. Without the coefficient, the value of second term can be much smaller than the first one so that the second term is neglected in the optimization.

The formulation above is an unconstrained optimization problem. Newton’s methods are widely used in solving this kind of problem. As it is computationally costly to calculate the inverse Hessian matrix in Newton’s method, Quasi-Newton methods are developed to approximate inverse Hessian matrix with less computation \cite{40}. BFGS method is one of the most effective and numerically stable methods in all the Quasi-Newton methods \cite{41}. As the dimensions of matrices can be large, we use L-BFGS to solve the problem. L-BFGS is modified based on BFGS method. Compared to BFGS, it only stores the last few gradient pairs to obtain the inverse Hessian approximation for saving storage space and computing resources \cite{42}.

The optimal solution obtained through L-BFGS is denoted as ${{\mathbf{s}}_{0}}$. It is important as a reference radar waveform implemented in the design of the ISAC waveform.
Note that ${{\mathbf{s}}_{0}}$ does not necessarily satisfy the energy budget. Therefore, it is important to perform energy normalization on the resulting ${{\mathbf{s}}_{0}}$, i.e.,
\begin{equation}
{{\mathbf{s}}_{0\_\text{norm}}}=\frac{{{\mathbf{s}}_{0}}}{{{\left\| {{\mathbf{s}}_{0}} \right\|}_{2}}}
\end{equation}

The ideal waveform mentioned below refers to the normalized one.

\subsubsection{Design of ideal radar waveform with oversampling}
\label{s3-1-2}
The transmission symbol vector of detection angle $\theta $ can be transformed into
\begin{equation}
\begin{aligned}
   {{{\mathbf{\tilde{s}}}}_{\text{v}}}(\theta )&={{[{{{\mathbf{\tilde{s}}}}^{\text{T}}}({{\mathbf{I}}_{\text{ }\!\!\gamma\!\!\text{ }{{\text{N}}_{\text{s}}}\text{+ }\!\!\gamma\!\!\text{ }{{\text{N}}_{\text{cp}}}}}\otimes \mathbf{a}(\theta ))]}^{\text{T}}} \\ 
 & ={{({{\mathbf{I}}_{\text{ }\!\!\gamma\!\!\text{ }{{\text{N}}_{\text{s}}}\text{+ }\!\!\gamma\!\!\text{ }{{\text{N}}_{\text{cp}}}}}\otimes \mathbf{a}(\theta ))}^{\text{T}}}\mathbf{\tilde{s}} \\ 
 & ={{({{\mathbf{I}}_{\text{ }\!\!\gamma\!\!\text{ }{{\text{N}}_{\text{s}}}\text{+ }\!\!\gamma\!\!\text{ }{{\text{N}}_{\text{cp}}}}}\otimes \mathbf{a}(\theta ))}^{\text{T}}}\mathbf{\tilde{\Gamma }}{{\mathbf{s}}_{\text{os}}} . 
\end{aligned}
\end{equation}

For simplicity of presentation, matrix $\mathbf{\tilde{G}}(\theta )$ is defined as 
\begin{equation}
\mathbf{\tilde{G}}(\theta )={{({{\mathbf{I}}_{\text{ }\!\!\gamma\!\!\text{ }{{\text{N}}_{\text{s}}}\text{+ }\!\!\gamma\!\!\text{ }{{\text{N}}_{\text{cp}}}}}\otimes \mathbf{a}(\theta ))}^{\text{T}}}\mathbf{\tilde{\Gamma }}.
\end{equation}
With $\mathbf{\tilde{G}}(\theta )$, the vector of transmission symbols in the direction of $\theta $ can be expressed as
\begin{equation}
	{{\mathbf{\tilde{s}}}_{\text{v}}}(\theta )=\mathbf{\tilde{G}}(\theta ){{\mathbf{s}}_{\text{os}}}.
\end{equation}

According to \eqref{ambiguity}, the summation of ambiguity function values to be minimized is
\begin{equation}
\sum\limits_{\theta \in {{\Omega }_{d}}}{\sum\limits_{(k,f)\in \Delta }{{{\left| \mathbf{s}_{\text{os}}^{\text{H}}{{{\mathbf{\tilde{G}}}}^{\text{H}}}(\theta ){{{\mathbf{\tilde{J}}}}_{\text{k}}}{{{\mathbf{\tilde{D}}}}_{\text{f}}}\mathbf{\tilde{G}}(\theta ){{\mathbf{s}}_{\text{os}}} \right|}^{2}}}}.
\end{equation}

The beam pattern with oversampling is
\begin{equation}
\frac{1}{\gamma {{N}_{\text{s}}}+\gamma {{N}_{\text{cp}}}}\mathbf{s}_{\text{os}}^{\text{H}}{{\mathbf{\tilde{G}}}^{\text{H}}}(\theta )\mathbf{\tilde{G}}(\theta ){{\mathbf{s}}_{\text{os}}},
\end{equation}
based on which the difference value between realistic and ideal beam patterns can be expressed as 
\begin{equation}
{{[\frac{1}{\gamma {{N}_{\text{s}}}+\gamma {{N}_{\text{cp}}}}\mathbf{s}_{\text{os}}^{\text{H}}{{\mathbf{\tilde{G}}}^{\text{H}}}(\theta )\mathbf{\tilde{G}}(\theta ){{\mathbf{s}}_{\text{os}}}-d(\theta )]}^{2}}.
\end{equation}

The design of ideal radar waveform with oversampling is 
\begin{small}
\begin{equation}
\begin{aligned}
  & \underset{{{\mathbf{s}}_{\text{os}}}}{\mathop{\min }}\,\sum\limits_{\theta \in {{\Omega }_{d}}}{\sum\limits_{(k,f)\in \Delta }{{{\left| \mathbf{s}_{\text{os}}^{\text{H}}{{{\mathbf{\tilde{G}}}}^{\text{H}}}(\theta ){{{\mathbf{\tilde{J}}}}_{\text{k}}}{{{\mathbf{\tilde{D}}}}_{\text{f}}}\mathbf{\tilde{G}}(\theta ){{\mathbf{s}}_{\text{os}}} \right|}^{2}}}}+ \\ 
 & {{(\gamma {{N}_{\text{s}}}+\gamma {{N}_{\text{cp}}})}^{2}}\sum\limits_{\theta \in {{\Omega }_{all}}}{{{[\frac{\mathbf{s}_{\text{os}}^{\text{H}}{{{\mathbf{\tilde{G}}}}^{\text{H}}}(\theta )\mathbf{\tilde{G}}(\theta ){{\mathbf{s}}_{\text{os}}}}{\gamma {{N}_{\text{s}}}+{{N}_{\text{cp}}}}-d(\theta )]}^{2}}} .\\ 
\end{aligned}
\end{equation}
\end{small}

This is also an optimization problem without constraints, which can be solved with L-BFGS. The obtained optimal solution is denoted as ${{\mathbf{\tilde{s}}}_{0}}$, which, after energy normalization, can be used in the design of ISAC waveform with oversampling as a reference.

\subsection{Design of ISAC MIMO-OFDM waveform}
\label{s3-2}

\subsubsection{Problem formulation with Nyquist sampling}
\label{s3-2-1}

In the optimization problem, we minimize the MUI to guarantee the quality of communication. Meanwhile, the gap between the beam pattern of designed waveform and ideal beam pattern is narrowed down. We also minimize the ambiguity function of interested region in the objective function while restraining PAPRs in the constraints. Then the objective function is formulated as
\begin{equation}
\begin{aligned}
  & \underset{\mathbf{x}}{\mathop{\min }}\,\frac{\rho }{\left\| {{\mathbf{s}}_{\text{D}}} \right\|_{2}^{2}}\left\| \mathbf{Hx}-{{\mathbf{s}}_{\text{D}}} \right\|_{2}^{2} \\ 
 & +\frac{1-\rho }{\left\| {{\mathbf{s}}_{0}} \right\|_{2}^{2}}\left\| ({{\mathbf{F}}^{\text{H}}}\otimes {{\mathbf{I}}_{{{\text{N}}_{\text{t}}}}})\mathbf{x}-{{\mathbf{s}}_{0}} \right\|_{2}^{2} ,\\ 
\end{aligned}
\end{equation}
where the former represents the communication multi-user interference and the latter is the similarity constraint on the waveform. Herein, ${{\mathbf{s}}_{0}}$ is an ideal radar waveform. To be noted that the beam pattern and ambiguity function are considered in the design of ${{\mathbf{s}}_{0}}$.

PAPR is the ratio of peak power to average power, so the PAPR constraint can be decomposed into two constraints: the constraint of total power and the power constraint of each symbol. 

Since the cyclic prefix symbols are the same as the last symbols of the original data, the power of CP is equal to $\frac{{{N}_{cp}}}{{{N}_{cp}}+{{N}_{s}}}$ of the total power. Therefore, the constraint on the total energy of all symbols in the time domain can be transformed into a constraint on the total energy of effective symbols, which is
\begin{equation}
\left\| {{{\mathbf{\hat{s}}}}_{\text{l}}} \right\|_{2}^{2}=\frac{{{N}_{\text{s}}}}{{{N}_{\text{cp}}}+{{N}_{\text{s}}}}\frac{{{E}_{\text{t}}}}{{{N}_{\text{t}}}}.
\end{equation}
The equation is equivalent to 
\begin{equation}
\left\| {{\mathbf{C}}_{\text{l}}}({{\mathbf{F}}^{\text{H}}}\otimes {{\mathbf{I}}_{{{\text{N}}_{\text{t}}}}})\mathbf{x} \right\|_{2}^{2}=\frac{{{N}_{\text{s}}}}{{{N}_{\text{cp}}}+{{N}_{\text{s}}}}\frac{{{E}_{\text{t}}}}{{{N}_{\text{t}}}},
\end{equation}
where ${{E}_{\text{t}}}$ represents the total energy available to the system corresponding to ${{N}_{\text{t}}}$ antennas and ${{N}_{\text{s}}}+{{N}_{\text{cp}}}$ OFDM symbols.

With a fixed total energy, the average power is determined so that the constraints of PAPRs can be transformed into the constraints of peak power. The constraints of peak power can be further converted to the power constraint of every symbol. Therefore, the constraints can be written as 
\begin{equation}
\begin{aligned}
  & {{\left| {{[{{\mathbf{C}}_{\text{l}}}({{\mathbf{F}}^{\text{H}}}\otimes {{\mathbf{I}}_{{{\text{N}}_{\text{t}}}}})\mathbf{x}]}_{\text{n}}} \right|}^{2}}\le {{\gamma }_{\text{l}}}, \\ 
 & l\in \left\{ 1,\cdots ,{{N}_{\text{t}}} \right\},n\in \left\{ 1,\cdots ,{{N}_{\text{s}}} \right\} .\\ 
\end{aligned}
\end{equation}

The design of ISAC waveform with Nyquist sampling is formulated as
\begin{equation}
\begin{aligned}
   \underset{\mathbf{x}}{\mathop{\min }}&\,\frac{\rho }{\left\| {{\mathbf{s}}_{\text{D}}} \right\|_{2}^{2}}\left\| \mathbf{Hx}-{{\mathbf{s}}_{\text{D}}} \right\|_{2}^{2} \\ 
 & +\frac{1-\rho }{\left\| {{\mathbf{s}}_{0}} \right\|_{2}^{2}}\left\| ({{\mathbf{F}}^{\text{H}}}\otimes {{\mathbf{I}}_{{{\text{N}}_{\text{t}}}}})\mathbf{x}-{{\mathbf{s}}_{0}} \right\|_{2}^{2} \\ 
  \text{s}\text{.t}\text{.}&{{\left| {{[{{\mathbf{C}}_{\text{l}}}({{\mathbf{F}}^{\text{H}}}\otimes {{\mathbf{I}}_{{{\text{N}}_{\text{t}}}}})\mathbf{x}]}_{\text{n}}} \right|}^{2}}\le {{\gamma }_{\text{l}}}, \\ 
 & l\in \left\{ 1,\cdots ,{{N}_{t}} \right\},n\in \left\{ 1,\cdots ,{{N}_{s}} \right\} \\ 
 & \left\| {{\mathbf{C}}_{\text{l}}}({{\mathbf{F}}^{\text{H}}}\otimes {{\mathbf{I}}_{{{\text{N}}_{\text{t}}}}})\mathbf{x} \right\|_{2}^{2}=\frac{{{N}_{\text{s}}}}{{{N}_{\text{cp}}}+{{N}_{\text{s}}}}\frac{{{E}_{\text{t}}}}{{{N}_{\text{t}}}}, \\ 
 & l\in \left\{ 1,\cdots ,{{N}_{\text{t}}} \right\} .\\ 
\end{aligned}
\label{nyquist}
\end{equation}

\subsubsection{Problem formulation with oversampling}
\label{s3-2-2}

When the oversampling rate is set to 1, oversampling degrates to Nyquist sampling. The case of Nyquist sampling can be regarded as a special case of oversampling. However, to make it easier for readers to understand, we formulate the optimization problem of Nyquist sampling independently before we discuss the case of oversampling which is more general. According to the definition of PAPRs with oversampling, the design of the ISAC waveform with oversampling is formulated as 
\begin{equation}
\begin{aligned}
   \underset{\mathbf{x}}{\mathop{\min }}&\,\frac{\rho }{\left\| {{\mathbf{s}}_{\text{D}}} \right\|_{2}^{2}}\left\| \mathbf{Hx}-{{\mathbf{s}}_{\text{D}}} \right\|_{2}^{2} \\ 
 & +\frac{1-\rho }{\left\| {{\mathbf{s}}_{0}} \right\|_{2}^{2}}\left\| (\mathbf{F}_{\text{os}}^{\text{H}}\otimes {{\mathbf{I}}_{{{\text{N}}_{\text{t}}}}})\mathbf{X}-{{{\mathbf{\tilde{s}}}}_{0}} \right\|_{2}^{2} \\ 
  \text{s}\text{.t}\text{.}&{{\left| {{[{{{\mathbf{\tilde{C}}}}_{\text{l}}}(\mathbf{\tilde{F}}_{\text{os}}^{\text{H}}\otimes {{\mathbf{I}}_{{{\text{N}}_{\text{t}}}}})\mathbf{x}]}_{\text{n}}} \right|}^{2}}\le {{\gamma }_{\text{l}}}, \\ 
 & l\in \left\{ 1,\cdots ,{{N}_{\text{t}}} \right\},n\in \left\{ 1,\cdots ,\gamma {{N}_{\text{s}}} \right\} \\ 
 & \left\| [{{{\mathbf{\tilde{C}}}}_{\text{l}}}(\mathbf{\tilde{F}}_{\text{os}}^{\text{H}}\otimes {{\mathbf{I}}_{{{\text{N}}_{\text{t}}}}})\mathbf{x}] \right\|_{2}^{2}=\frac{\gamma {{N}_{\text{s}}}}{{{N}_{\text{cp}}}+\gamma {{N}_{\text{s}}}}\frac{{{E}_{\text{t}}}}{{{N}_{\text{t}}}}, \\ 
 & l\in \left\{ 1,\cdots ,{{N}_{\text{t}}} \right\} .\\ 
\end{aligned}
\label{oversampling}
\end{equation}

\subsubsection{ADMM for solving optimization problem}
\label{s3-2-3}

The optimization problem of ISAC MIMO-OFDM waveform is \eqref{nyquist} for Nyquist sampling rate and \eqref{oversampling} for oversampling. It is obvious that the two optimization problems are in the same form. For conciseness, we only discuss the solving process of the optimization with Nyquist sampling rate. The problem with oversampling can be solved by the exact same method only with little modification.

It is obvious that $\mathbf{s}=({{\mathbf{F}}^{\text{H}}}\otimes {{\mathbf{I}}_{{{\text{N}}_{\text{t}}}}})\mathbf{x}$. Because $\mathbf{x}$ and $\mathbf{s}$ can be derived from each other, obtaining the optimal $\mathbf{s}$ is equivalent to obtaining the optimal $\mathbf{x}$. The subproblem can be further transformed to
\begin{equation}
\begin{aligned}
  \underset{\mathbf{s}}{\mathop{\min }} &\,\frac{\rho }{\left\| {{\mathbf{s}}_{\text{D}}} \right\|_{2}^{2}}\left\| \mathbf{H}(\mathbf{F}\otimes {{\mathbf{I}}_{{{\text{N}}_{\text{t}}}}})\mathbf{s}-{{\mathbf{s}}_{\text{D}}} \right\|_{2}^{2}+ \\ 
 & \frac{1-\rho }{\left\| {{\mathbf{s}}_{0}} \right\|_{2}^{2}}\left\| \mathbf{s}-{{\mathbf{s}}_{0}} \right\|_{2}^{2} \\ 
  \text{s}\text{.t}\text{.}&{{\left| {{[{{\mathbf{C}}_{\text{l}}}\mathbf{s}]}_{\text{n}}} \right|}^{2}}\le \varepsilon , \\ 
 & l\in \left\{ 1,\cdots ,{{N}_{t}} \right\},n\in \left\{ 1,\cdots ,{{N}_{s}} \right\} \\ 
 & \left\| {{\mathbf{C}}_{\text{l}}}\mathbf{s} \right\|_{2}^{2}=\frac{{{N}_{s}}}{{{N}_{cp}}+{{N}_{s}}}\frac{{{E}_{t}}}{{{N}_{t}}}, \\ 
 & l\in \left\{ 1,\cdots ,{{N}_{t}} \right\} .\\ 
\end{aligned}
\end{equation}

For simplicity, define $\mathbf{\hat{H}}=\mathbf{H}(\mathbf{F}\otimes {{\mathbf{I}}_{{{\text{N}}_{\text{t}}}}})$. The problem can be rewritten into a more concise form as

\begin{equation}
\begin{aligned}
   \underset{\mathbf{s}}{\mathop{\min }}&\,\frac{\rho }{\left\| {{\mathbf{s}}_{\text{D}}} \right\|_{2}^{2}}\left\| \mathbf{\hat{H}s}-{{\mathbf{s}}_{\text{D}}} \right\|_{2}^{2} \\
   &+ \frac{1-\rho }{\left\| {{\mathbf{s}}_{0}} \right\|_{2}^{2}}\left\| \mathbf{s}-{{\mathbf{s}}_{0}} \right\|_{2}^{2} \\ 
  \text{s}\text{.t}\text{.}&{{\left| {{[{{\mathbf{C}}_{\text{l}}}\mathbf{s}]}_{\text{n}}} \right|}^{2}}\le \varepsilon , \\ 
 & l\in \left\{ 1,\cdots ,{{N}_{\text{t}}} \right\},n\in \left\{ 1,\cdots ,{{N}_{\text{s}}} \right\} \\ 
 & \left\| {{\mathbf{C}}_{\text{l}}}\mathbf{s} \right\|_{2}^{2}=\frac{{{N}_{\text{s}}}}{{{N}_{\text{cp}}}+{{N}_{\text{s}}}}\frac{{{E}_{\text{t}}}}{{{N}_{\text{t}}}}, \\ 
 & l\in \left\{ 1,\cdots ,{{N}_{\text{t}}} \right\} .\\ 
\end{aligned}
 \label{optimization_problem}
\end{equation}

Note that problem \eqref{optimization_problem} is still a non-convex QCQP problem which is hard to solve directly. Although \cite{30} has proposed an SDR-based algorithm to solve a problem similar to \eqref{optimization_problem}. However, the PAPR and power constraints used in \cite{30} are quite different from ours, and hence there is no rank-one guarantee for our problem if the SDR-based approach applies to \eqref{optimization_problem}. In addition, the SDR-based approach will significantly increase the computational cost. To this end, we propose an efficient algorithm based on ADMM to solve \eqref{optimization_problem}. Note that \eqref{optimization_problem} has two classes of constraints which are coupled with each other. To address this issue, we introduce two auxilliary vectors $\mathbf{y}$ and $\mathbf{v}$. By doing so, the original problem \eqref{optimization_problem} can be equivalently transformed to
\begin{small}
\begin{equation}
\begin{aligned}
   \underset{\mathbf{s},\mathbf{y},\mathbf{v}}{\mathop{\min }}&\,\frac{\rho }{\left\| {{\mathbf{s}}_{\text{D}}} \right\|_{2}^{2}}\left\| \mathbf{\hat{H}s}-{{\mathbf{s}}_{\text{D}}} \right\|_{2}^{2}+\frac{1-\rho }{\left\| {{\mathbf{s}}_{0}} \right\|_{2}^{2}}\left\| \mathbf{s}-{{\mathbf{s}}_{0}} \right\|_{2}^{2} \\ 
  \text{s}\text{.t}\text{.}&\mathbf{y}=\mathbf{s} \\ 
 & \mathbf{v}=\mathbf{s} \\ 
 & {{\left| {{[{{\mathbf{C}}_{\text{l}}}\mathbf{y}]}_{\text{n}}} \right|}^{2}}\le \varepsilon , \\ 
 & l\in \left\{ 1,\cdots ,{{N}_{\text{t}}} \right\},n\in \left\{ 1,\cdots ,{{N}_{\text{s}}} \right\} \\ 
 & \left\| {{\mathbf{C}}_{\text{l}}}\mathbf{v} \right\|_{2}^{2}=\frac{{{N}_{\text{s}}}}{{{N}_{\text{cp}}}+{{N}_{\text{s}}}}\frac{{{E}_{\text{t}}}}{{{N}_{\text{t}}}}, \\ 
 & l\in \left\{ 1,\cdots ,{{N}_{\text{t}}} \right\} .\\ 
\end{aligned}
\end{equation}
\end{small}

The corresponding augmented Lagrange function is 
\begin{equation}
\begin{aligned}
   &{{L}_{\eta }}(\mathbf{s},\mathbf{y},\mathbf{\lambda },\mathbf{v},\mathbf{\mu })=\frac{\rho }{\left\| {{\mathbf{s}}_{\text{D}}} \right\|_{2}^{2}}\left\| \mathbf{\hat{H}s}-{{\mathbf{s}}_{\text{D}}} \right\|_{2}^{2}+ \\ 
 & \frac{1-\rho }{\left\| {{\mathbf{s}}_{0}} \right\|_{2}^{2}}\left\| \mathbf{s}-{{\mathbf{s}}_{0}} \right\|_{2}^{2}+  
 \frac{\eta }{2}({{\left\| \mathbf{y}-\mathbf{s}+\mathbf{\lambda } \right\|}_{2}^{2}}-{{\left\| \mathbf{\lambda } \right\|}_{2}^{2}})+ \\ 
 & \frac{\eta }{2}({{\left\| \mathbf{v}-\mathbf{s}+\mathbf{\mu } \right\|}_{2}^{2}}-{{\left\| \mathbf{\mu } \right\|}_{2}^{2}}), \\ 
\end{aligned}
\label{lagrange function}
\end{equation}
where $\mathbf{\lambda }$ and $\mathbf{\mu }$ are Lagrange multipliers. The specific procedure of solving the problem by ADMM is presented as follows.

\textbf{Step 1:} Updating $\{\mathbf{y},\mathbf{v}\}$

According to the principle of ADMM, the update of $\{\mathbf{y},\mathbf{v}\}$ is 
%\begin{strip}

\begin{equation}\label{eq78}
\begin{array}{l}
\{ {{\bf{y}}^{(m + 1)}},{{\bf{v}}^{(m + 1)}}\}  = \\
\mathop {\arg \min }\limits_{\begin{array}{*{20}{c}}
{{{\left| {{{[{{\bf{C}}_\text{l}}{\bf{y}}]}_n}} \right|}^2} \le \varepsilon ,l \in \left\{ {1, \cdots ,{N_t}} \right\},n \in \left\{ {1, \cdots ,{N_s}} \right\}}\\
{\left\| {{{\bf{C}}_\text{l}}{\bf{v}}} \right\|_2^2 = \frac{{{N_\text{s}}}}{{{N_{cp}} + {N_s}}}\frac{{{E_t}}}{{{N_t}}},l \in \left\{ {1, \cdots ,{N_\text{t}}} \right\}}
\end{array}} {\mkern 1mu} \\\\
\;\;\;\;\;\;\;\;\;\;\;{L_\eta }({{\bf{s}}^{(m)}},{\bf{y}},{{\bf{\lambda }}^{(m)}},{\bf{v}},{{\bf{\mu }}^{(m)}}).
\end{array}
\end{equation}
%\end{strip}%

When solving ${{\mathbf{y}}^{(m+1)}}$, the minimization problem can be formulated specifically as
\begin{equation}
\begin{aligned}
   \underset{\mathbf{y}}{\mathop{\min }}&\,{{\left\| \mathbf{y}-\mathbf{s}^{(m)}+{{\mathbf{\lambda }}^{(m)}} \right\|}_{2}^{2}} \\ 
  \text{s}\text{.t}\text{.}&{{\left| {{[{{\mathbf{C}}_{\text{l}}}\mathbf{y}]}_{\text{n}}} \right|}^{2}}\le \varepsilon , \\ 
 & l\in \left\{ 1,\cdots ,{{N}_{\text{t}}} \right\},n\in \left\{ 1,\cdots ,{{N}_{\text{s}}} \right\} ,\\ 
\end{aligned}
\end{equation}
where ${{\mathbf{C}}_{\text{l}}}$ is a choosing matrix, whose elements are 0 or 1. The choosing matrix  ${{\mathbf{C}}_{\text{l}}}$ selects the elements corresponding to the \textit{l}-th antenna and combines them into a new vector. $\left[ {{\mathbf{C}}_{\text{l}}}\mathbf{y} \right]$ contains the ${{\mathbf{N}}_{s}}$ symbols of the \textit{l}-th antenna, with ${{\left[ {{\mathbf{C}}_{\text{l}}}\mathbf{y} \right]}_{\text{n}}}$ representing the \textit{n}-th symbol. Because the constraint covers all the $l\in \left\{ 1,\cdots ,{{N}_{\text{t}}} \right\}$ and $n\in \left\{ 1,\cdots ,{{N}_{\text{s}}} \right\}$, the minimization problem can be transformed equivalently to
\begin{equation}
\begin{aligned}
  & \underset{\mathbf{y}}{\mathop{\min }}\,{{\left\| \mathbf{y}-{{\mathbf{s}}^{(m)}}+{{\mathbf{\lambda }}^{(m)}} \right\|}_{2}^{2}} \\ 
 & \text{s}\text{.t}\text{.}{{\left| {{[\mathbf{y}]}_{\text{i}}} \right|}^{2}}\le \varepsilon ,i\in \left\{ 1,\cdots ,{{N}_{\text{s}}}{{N}_{\text{t}}} \right\} .\\ 
\end{aligned}
\label{admm_y}
\end{equation}

It can be decomposed further into ${{N}_{\text{s}}}{{N}_{\text{t}}}$ subproblems with $i\in \left\{ 1,\cdots ,{{N}_{s}}{{N}_{t}} \right\}$. The subproblem is formulated as
\begin{equation}
\begin{aligned}
  & \underset{{{\left[\mathbf{y}\right]}_{\text{i}}}}{\mathop{\min }}\,{{\left| {{\left[\mathbf{y}\right]}_{\text{i}}}-{{\left[{{\mathbf{s}}^{(m)}}\right]}_{\text{i}}}+{{\left[{{\mathbf{\lambda }}^{(m)}}\right]}_{\text{i}}} \right|}^{2}} \\ 
 & \text{s}\text{.t}\text{.}{{\left| {{\left[\mathbf{y}\right]}_{\text{i}}} \right|}^{2}}\le \varepsilon . \\ 
\end{aligned}
\end{equation}

It is easy to prove that the solution of the subproblem is
\begin{equation}
\begin{aligned}
  & [\mathbf{y}]_{\text{i}}^{(m+1)}= \\ 
 & \left\{ \begin{matrix}
   {{[{{\mathbf{s}}^{(m)}}]}_{\text{i}}}-{{[{{\mathbf{\lambda }}^{(m)}}]}_{\text{i}}} & {{\left| {{[{{\mathbf{s}}^{(m)}}]}_{\text{i}}}-{{[{{\mathbf{\lambda }}^{(m)}}]}_{\text{i}}} \right|}^{2}}\le \varepsilon   \\
   \sqrt{\varepsilon }\frac{{{[{{\mathbf{s}}^{(m)}}]}_{\text{i}}}-{{[{{\mathbf{\lambda }}^{(m)}}]}_{\text{i}}}}{\left| {{[{{\mathbf{s}}^{(m)}}]}_{\text{i}}}-{{[{{\mathbf{\lambda }}^{(m)}}]}_{\text{i}}} \right|} & \text{otherwise}  \\
\end{matrix} \right. .\\ 
\end{aligned}
\label{y_ADMM}
\end{equation}

The value of $\mathbf{v}$ can be updated by the following minimization problem,i.e.,
\begin{equation}
\begin{aligned}
   \underset{\mathbf{v}}{\mathop{\min }}&\,{{\left\| \mathbf{v}-{{\mathbf{s}}^{(m)}}+{{\mathbf{\mu }}^{(m)}} \right\|}_{2}^{2}} \\ 
 \text{s}\text{.t}\text{.} &\left\| {{\mathbf{C}}_{\text{l}}}\mathbf{v} \right\|_{\text{F}}^{2}=\frac{{{N}_{\text{s}}}}{{{N}_{\text{cp}}}+{{N}_{\text{s}}}}\frac{{{E}_{\text{t}}}}{{{N}_{\text{t}}}}, \\ 
 &l\in \left\{ 1,\cdots ,{{N}_{\text{t}}} \right\} .\\ 
\end{aligned}
\end{equation}

As the constraint covers all the $l\in \left\{ 1,\cdots ,{{N}_{\text{t}}} \right\}$, the problem can be decomposed into ${{N}_{\text{t}}}$ subproblems. The \textit{l}-th subproblem is
\begin{equation}
\begin{aligned}
  & \underset{{{{\mathbf{\tilde{v}}}}_{\text{l}}}}{\mathop{\min }}\,{{\left\| {{{\mathbf{\tilde{v}}}}_{\text{l}}}-{{\mathbf{C}}_{\text{l}}}({{\mathbf{s}}^{(m)}}-{{\mathbf{\mu }}^{(m)}}) \right\|}_{2}^{2}} \\ 
 & \text{s}\text{.t}\text{.}\left\| {{{\mathbf{\tilde{v}}}}_{\text{l}}} \right\|_{\text{F}}^{2}=\frac{{{N}_{\text{s}}}}{{{N}_{\text{cp}}}+{{N}_{\text{s}}}}\frac{{{E}_{\text{t}}}}{{{N}_{\text{t}}}} ,\\ 
\end{aligned}
\end{equation}
where ${{\mathbf{\tilde{v}}}_{\text{l}}}={{\mathbf{C}}_{\text{l}}}\mathbf{v}$. It can be easily proved that the optimal solution is
\begin{small}
\begin{equation}
\!\mathbf{\tilde{v}}_{\text{l}}^{(m+1)}=\sqrt{\frac{{{N}_{\text{s}}}}{{{N}_{\text{cp}}}+{{N}_{\text{s}}}}\frac{{{E}_{\text{t}}}}{{{N}_{\text{t}}}}}\frac{{{\mathbf{C}}_{\text{l}}}({{\mathbf{s}}^{(m)}}-{{\mathbf{\mu }}^{(m)}})}{\left\| {{\mathbf{C}}_{\text{l}}}({{\mathbf{s}}^{(m)}}-{{\mathbf{\mu }}^{(m)}}) \right\|}_{2}.\!
\label{v_ADMM}
\end{equation}
\end{small}

After calculating all the ${{\mathbf{\tilde{v}}}_{\text{l}}}$ with $l\in \left\{ 1,\cdots ,{{N}_{\text{t}}} \right\}$, we rearrange the elements of these ${{N}_{\text{t}}}$ vectors into one vector. As the ${{N}_{\text{t}}}$ vectors contain all the elements of $\mathbf{v}$ in a regular form, the vector rearranged from them is $\mathbf{v}$.

\textbf{Step 2:} Updating $\mathbf{s}$

$\mathbf{s}$ is updated by solving an optimization problem with given ${\mathbf{y}}^{(m+1)}$, ${\mathbf{\lambda }}^{(m)}$, ${\mathbf{v}}^{(m+1)}$ and ${\mathbf{\mu }}^{(m)}$, i.e.,
\begin{small}
\begin{equation}
\begin{aligned}
	&{{\mathbf{s}}^{(m+1)}}=\\
 &\underset{\mathbf{s}}{\mathop{\arg \min }}\,{{L}_{\eta }}(\mathbf{s},{{\mathbf{y}}^{(m+1)}},{{\mathbf{\lambda }}^{(m)}},{{\mathbf{v}}^{(m+1)}},{{\mathbf{\mu }}^{(m)}}).
 \end{aligned}
\end{equation}
\end{small}

Based on the augmented Lagrange function, the problem can be expressed specifically as

\begin{equation}
\begin{aligned}
 \! & \underset{\mathbf{s}}{\mathop{\min }}\,\frac{\rho }{\left\| {{\mathbf{s}}_{\text{D}}} \right\|_{2}^{2}}\left\| \mathbf{\hat{H}s}-{{\mathbf{s}}_{\text{D}}} \right\|_{2}^{2}+\frac{1-\rho }{\left\| {{\mathbf{s}}_{0}} \right\|_{2}^{2}}\left\| \mathbf{s}-{{\mathbf{s}}_{0}} \right\|_{2}^{2} \\ 
 & +\frac{\eta }{2}{{\left\| {{\mathbf{y}}^{(m+1)}}-\mathbf{s}+{{\mathbf{\lambda }}^{(m)}} \right\|}_{2}^{2}}\\
 &+\frac{\eta }{2}{{\left\| {{\mathbf{v}}^{(m+1)}}-\mathbf{s}+{{\mathbf{\mu }}^{(m)}} \right\|}_{2}^{2}}\! , 
\end{aligned}
\end{equation}

which is an unconstrained quadratic optimization problem. Setting the derivative to zero and simplifying the equation, the optimal $\mathbf{s}$ can be acquired by solving the equation of
\begin{equation}
	\mathbf{As}=\mathbf{b},
\end{equation}
where $\mathbf{A}$ and $\mathbf{b}$ are defined as
\begin{equation}
\mathbf{A}=\frac{\rho }{\left\| {{\mathbf{s}}_{\text{D}}} \right\|_{2}^{2}}{{\mathbf{\hat{H}}}^{\text{H}}}\mathbf{\hat{H}}+(\frac{1-\rho }{\left\| {{\mathbf{s}}_{0}} \right\|_{2}^{2}}+\eta ){{\mathbf{I}}_{{{\text{N}}_{\text{s}}}{{\text{N}}_{\text{t}}}}}
\end{equation}
and
\begin{equation}
\begin{aligned}
  & \mathbf{b}=\frac{\rho }{\left\| {{\mathbf{s}}_{\text{D}}} \right\|_{2}^{2}}{{{\mathbf{\hat{H}}}}^{\text{H}}}{{\mathbf{s}}_{\text{D}}}+\frac{1-\rho }{\left\| {{\mathbf{s}}_{0}} \right\|_{2}^{2}}{{\mathbf{s}}_{0}} \\ 
 & +\frac{\eta }{2}({{\mathbf{y}}^{(m+1)}}+{{\mathbf{\lambda }}^{(m)}}+{{\mathbf{v}}^{(m+1)}}+{{\mathbf{\mu }}^{(m)}}) .\\ 
\end{aligned}
\end{equation}

The least square solution is used as the iterative value of $\mathbf{s}$ in the (\textit{m}+1)-th  iteration:
\begin{equation}
	{{\mathbf{s}}^{(m+1)}}={{({{\mathbf{A}}^{\text{H}}}\mathbf{A})}^{-1}}{{\mathbf{A}}^{\text{H}}}\mathbf{b}.
 \label{S_ADMM}
\end{equation}

\textbf{Step 3:} Updating $\{\mathbf{\lambda },\mathbf{\mu }\}$

The Lagrange multipliers $\mathbf{\lambda }$ and $\mathbf{\mu }$ are updated as
\begin{equation}
	{{\mathbf{\lambda }}^{(m+1)}}={{\mathbf{\lambda }}^{(m)}}+{{\mathbf{y}}^{(m+1)}}-{{\mathbf{s}}^{(m+1)}}
 \label{lemda_ADMM}
\end{equation}
and
\begin{equation}
	{{\mathbf{\mu }}^{(m+1)}}={{\mathbf{\mu }}^{(m)}}+{{\mathbf{v}}^{(m+1)}}-{{\mathbf{s}}^{(m+1)}}.
 \label{miu_ADMM}
\end{equation}

\begin{algorithm}[H]
\caption{Proposed ADMM algorithm.}\label{algorithm1}
\begin{algorithmic}
        \STATE {\textbf{Require} $\rho $, ${{\mathbf{s}}_{\text{D}}}$, ${\mathbf{\hat{H}}}$, ${{\mathbf{s}}_{0}}$, $\eta$}

        \STATE {\textbf{Initialize:} ${{\mathbf{s}}^{(0)}}$, ${{\mathbf{y}}^{(0)}}$, ${{\mathbf{\lambda }}^{(0)}}$, ${{\mathbf{v}}^{(0)}}$, ${{\mathbf{\mu }}^{(0)}}$}
        \WHILE{not converged}
        \STATE {$m \gets m+1$} 
        \STATE {Update ${{\mathbf{y}}^{(m+1)}}$ and ${{\mathbf{v}}^{(m+1)}}$ via \eqref{y_ADMM} and \eqref{v_ADMM};}
        \STATE {Update ${{\mathbf{s}}^{(m+1)}}$ via \eqref{S_ADMM};}
        \STATE {Update ${{\mathbf{\lambda }}^{(m+1)}}$ and ${{\mathbf{\mu }}^{(m+1)}}$ via \eqref{lemda_ADMM} and \eqref{miu_ADMM};}
        \ENDWHILE
        \ENSURE ${{\mathbf{s}}^{(m+1)}}$
\end{algorithmic}
\end{algorithm}

\subsection{Convergence analysis of ADMM}
\label{s3-3}

    With the last two terms combined, the augmented Lagrange function \eqref{lagrange function} can be rewritten as
    \begin{equation}
       \begin{aligned}
       	 & {{L}_{\eta }}\left( \mathbf{s},\mathbf{y}\text{,}\mathbf{\lambda }\text{,}\mathbf{v}\text{,}\mathbf{\mu } \right)
       =\frac{\rho }{\left\| {{\mathbf{s}}_{\text{D}}} \right\|_{2}^{2}}\left\| \mathbf{\hat{H}s}-{{\mathbf{s}}_{\text{D}}} \right\|_{2}^{2}\\
        &+\frac{1-\rho }{\left\| {{\mathbf{s}}_{0}} \right\|_{2}^{2}}\left\| \mathbf{s}-{{\mathbf{s}}_{0}} \right\|_{2}^{2} \\ 
       	& +\frac{\eta }{2}\left( {{\left\| \left[ \begin{matrix}
       				\mathbf{y}  \\
       				\mathbf{v}  \\
       			\end{matrix} \right]-\left[ \begin{matrix}
       					\mathbf{s}  \\
       					\mathbf{s}  \\
       			\end{matrix} \right]+\left[ \begin{matrix}
       					\mathbf{\lambda }  \\
       					\mathbf{\mu }  \\
       			\end{matrix} \right] \right\|}_{2}^{2}}-{{\left\| \left[ \begin{matrix}
       				\mathbf{\lambda }  \\
       				\mathbf{\mu }  \\
       		\end{matrix} \right] \right\|}_{2}^{2}} \right).  
       \end{aligned}
    \end{equation}
For simplicity of presentation, the objective function $f\left( \mathbf{s} \right)$ can be represented as $\left\| \mathbf{Qs}-\mathbf{\beta } \right\|_{2}^{2}$ with $\mathbf{Q}={{\left[ \frac{\sqrt{\rho }}{{{\left\| {{\mathbf{s}}_\text{D}} \right\|}_{2}}}{{{\mathbf{\hat{H}}}}^{\text{T}}},\frac{\sqrt{1-\rho }}{{{\left\| {{\mathbf{s}}_{0}} \right\|}_{2}}}{{\mathbf{I}}_{{{N}_{s}}{{N}_{t}}}} \right]}^{\text{T}}}$ and $\mathbf{\beta }={{\left[ \frac{\sqrt{\rho }}{{{\left\| {{\mathbf{s}}_\text{D}} \right\|}_{2}}}\mathbf{s}_\text{D}^{\text{T}},\frac{\sqrt{1-\rho }}{{{\left\| {{\mathbf{s}}_{0}} \right\|}_{2}}}\mathbf{s}_{0}^{\text{T}} \right]}^{\text{T}}}$. We denote the $\left[ \begin{matrix}
       \mathbf{y}  \\
       \mathbf{v}  \\
    \end{matrix} \right]$ as $\mathbf{\xi }$ and $\left[ \begin{matrix}
    \mathbf{\lambda }  \\
       \mathbf{\mu }  \\
       \end{matrix} \right]$ as $\mathbf{\tilde{\lambda }}$.  Meanwhile, $\left[ \begin{matrix}
       \mathbf{s}  \\
       \mathbf{s}  \\
       \end{matrix} \right]$ can be expressed as $\mathbf{Ts}$ with $\mathbf{T}=\left[ {{\mathbf{I}}_{{{N}_{s}}{{N}_{t}}}};{{\mathbf{I}}_{{{N}_{s}}{{N}_{t}}}} \right]$. Therefore, the Lagrange function can be expressed equivalently as
       \begin{equation}
       	{{L}_{\eta }}\left( \mathbf{s},\mathbf{\xi }\text{,}\mathbf{\tilde{\lambda }} \right)=f\left( \mathbf{s} \right)+\frac{\eta }{2}\left( {{\left\| \mathbf{\xi }-\mathbf{Ts}+\mathbf{\tilde{\lambda }} \right\|}_{2}^{2}}-{{\left\| {\mathbf{\tilde{\lambda }}} \right\|}_{2}^{2}} \right).
       \end{equation}
       
       The  $\left( \text{m}+1 \right)$-th iteration of ADMM is as follows:
       \begin{equation}
       {{\mathbf{\xi }}^{\left( m+1 \right)}}=\underset{\mathbf{\xi }}{\mathop{\arg \min }}\,{{L}_{\eta }}\left( {{\mathbf{s}}^{\left( m \right)}},\mathbf{\xi },{{{\mathbf{\tilde{\lambda }}}}^{\left( m \right)}} \right)
       \end{equation}
       \begin{equation}
       {{\mathbf{s}}^{\left( m+1 \right)}}=\underset{\mathbf{s}}{\mathop{\arg \min }}\,{{L}_{\eta }}\left( \mathbf{s},{{\mathbf{\xi }}^{\left( m+1 \right)}},{{{\mathbf{\tilde{\lambda }}}}^{\left( m \right)}} \right)
       \end{equation}
       \begin{equation}
       {{\mathbf{\tilde{\lambda }}}^{\left( m+1 \right)}}={{\mathbf{\tilde{\lambda }}}^{\left( m \right)}}+{{\mathbf{\xi }}^{\left( m+1 \right)}}-\mathbf{T}{{\mathbf{s}}^{\left( m+1 \right)}}
       \end{equation}

       {\bf{Proposition 1:}}
        \label{proposition1}
        Assume that for any $\text{m}$, ${{\mathbf{\tilde{\lambda }}}^{\left( \text{m} \right)}}$ and ${{\mathbf{s}}^{\left( \text{m} \right)}}$ are bounded. We can find an $\varepsilon $ and an $\alpha $ to guarantee that $\left\| {{{\mathbf{\tilde{\lambda }}}}^{\left( m+1 \right)}}-{{{\mathbf{\tilde{\lambda }}}}^{\left( m \right)}} \right\|_{2}^{2}\le \frac{{{\alpha }^{2}}L_{f}^{2}}{{{\eta }^{2}}{{\varepsilon }^{2}}}\left\| {{\mathbf{s}}^{\left( m+1 \right)}}-{{\mathbf{s}}^{\left( m \right)}} \right\|_{2}^{2}$ . 
        
       {\bf{Proof:}}
        \label{proof1}
        See Appendix A.

       {\bf{Proposition 2:}}
        \label{proposition2}
        If $\eta \ge \frac{{{\alpha }^{2}}L_{f}^{2}}{{{\varepsilon }^{2}}{{\lambda }_{\min }}\left( {{\mathbf{Q}}^{T}}\mathbf{Q} \right)}$, ${{L}_{\eta }}\left( {{\mathbf{s}}^{\left( m \right)}},{{\mathbf{\xi }}^{\left( m \right)}}\text{,}{{{\mathbf{\tilde{\lambda }}}}^{\left( m \right)}} \right)$ is a non-increasing sequence.
        
      {\bf{Proof:}}
        \label{proof2}
        See Appendix B.

        {\bf{Proposition 3:}}
        \label{proposition3}
     ${{L}_{\eta }}\left( {{\mathbf{s}}^{\left( m \right)}},{{\mathbf{\xi }}^{\left( m \right)}}\text{,}{{{\mathbf{\tilde{\lambda }}}}^{\left( m \right)}} \right)$ has a lower bound.
     
     {\bf{Proof:}}
        \label{proof3}
        See Appendix C.

       Iterating the equation \eqref{eq21} gives 
       \begin{equation}
       	\begin{aligned}
       		& \left( {{\lambda }_{\min }}\left( {{\mathbf{Q}}^{T}}\mathbf{Q} \right)-\frac{{{\alpha }^{2}}L_{f}^{2}}{\eta {{\varepsilon }^{2}}} \right)\sum\limits_{m=0}^{\infty }{\left\| {{\mathbf{s}}^{\left( m+1 \right)}}-{{\mathbf{s}}^{\left( m \right)}} \right\|_{2}^{2}} \\ 
       		& \le {{L}_{\eta }}\left( {{\mathbf{s}}^{\left( 0 \right)}},{{\mathbf{\xi }}^{\left( 0 \right)}}\text{,}{{{\mathbf{\tilde{\lambda }}}}^{\left( 0 \right)}} \right)-~{{L}_{\eta }}\left( {{\mathbf{s}}^{\left( \infty  \right)}},{{\mathbf{\xi }}^{\left( \infty  \right)}}\text{,}{{{\mathbf{\tilde{\lambda }}}}^{\left( \infty  \right)}} \right) \\ 
       		& \le {{L}_{\eta }}\left( {{\mathbf{s}}^{\left( 0 \right)}},{{\mathbf{\xi }}^{\left( 0 \right)}}\text{,}{{{\mathbf{\tilde{\lambda }}}}^{\left( 0 \right)}} \right)-~{{b}_{L}} .\\ 
       	\end{aligned}
       \end{equation}
       
       If $\eta \ge \frac{{{\alpha }^{2}}L_{f}^{2}}{{{\varepsilon }^{2}}{{\lambda }_{\min }}\left( {{\mathbf{Q}}^{T}}\mathbf{Q} \right)}$, the inequality above implies that $\underset{m\to \infty }{\mathop{\lim }}\,{{\mathbf{s}}^{\left( m+1 \right)}}-{{\mathbf{s}}^{\left( m \right)}}=0$. According to ${{\left\| \nabla f\left( {{\mathbf{s}}^{\left( m+1 \right)}} \right)-\nabla f\left( {{\mathbf{s}}^{\left( m \right)}} \right) \right\|}_{2}}\le {{L}_{f}}{{\left\| {{\mathbf{s}}^{\left( m+1 \right)}}-{{\mathbf{s}}^{\left( m \right)}} \right\|}_{2}}$ and $\nabla f\left( {{\mathbf{s}}^{\left( m+1 \right)}} \right)=\eta {{\mathbf{T}}^{T}}{{\mathbf{\tilde{\lambda }}}^{\left( m+1 \right)}}$, it can be inferred that  $\underset{m\to \infty }{\mathop{\lim }}\,{{\mathbf{\tilde{\lambda }}}^{\left( m+1 \right)}}-{{\mathbf{\tilde{\lambda }}}^{\left( m \right)}}=\mathbf{0}$. With  ${{\mathbf{\tilde{\lambda }}}^{\left( m+1 \right)}}-{{\mathbf{\tilde{\lambda }}}^{\left( m \right)}}={{\mathbf{\xi }}^{\left( m+1 \right)}}-\mathbf{T}{{\mathbf{s}}^{\left( m+1 \right)}}$, we have $\underset{m\to \infty }{\mathop{\lim }}\,{{\mathbf{\xi }}^{\left( m \right)}}-\mathbf{T}{{\mathbf{s}}^{\left( m \right)}}=\mathbf{0}$.

\subsection{Computational complexity of the algorithm}
\label{s3-4}

%We will analyze the computational complexity of the algorithm in this subsection.

First, we need to generate the ideal radar waveform. The complexity of L-BFGS used in the design of the ideal radar waveform is ${\mathrm O}(N_{s}^{2}N_{t}^{2})$ with Nyquist sampling rate and ${\mathrm O}({{\gamma }^{2}}N_{s}^{2}N_{t}^{2})$ with oversampling. It is worth noting that the ideal radar waveform can be designed offline and saved in advance, which means the computational complexity of L-BFGS should not be contained in the complexity of real-time operation.

Then we optimize the final ISAC waveform by introducing the generated ideal radar waveform. The complexity of ADMM of each iteration can be calculated according to separate updates of the parameters. The complexity of updating $\mathbf{s}$ is ${\mathrm O}(2N_{s}^{3}N_{t}^{3}+N_{s}^{2}N_{t}^{2})$. The computational complexity of updating $\mathbf{y}$ is ${\mathrm O}({{N}_{s}}{{N}_{t}})$, which is the same as updating $\mathbf{v}$. When it comes to the complexity of updating auxiliary variables $\{\mathbf{\lambda },\mathbf{\mu }\}$, it is ${\mathrm O}({{N}_{s}}{{N}_{t}})$ for each of the parameter.

\section{Numerical simulation}
\label{s4}

In this section, we present simulation results to validate the effectiveness of the proposed waveform design approach under the circumstances of Nyquist-rate sampling and oversampling. The performance metrics of waveforms designed with different tradeoff between radar and communication, which is realized by changing $\rho $ in the design, are analyzed. This tradeoff demonstrates the variation of both radar performance and communication performance with the priority of design transferring from radar to communication. We also analyze the performance of waveforms designed under different PAPR constraints to reveal the effect of PAPR constraints on radar and communication performance.

Based on the scenario of V2X, we set the simulation parameters listed in Table \ref{tab1}. We assume the communication channel to be Rician fading channel whose principal component direction is consistent with the directions of radar targets. The channel is constructed with 4 non-zero channel taps and the operating frequency of the system is 28 GHz. The basic waveform of MIMO-OFDM is applied with 40 effective OFDM symbols in time domain. The subcarrier interval is set as 300 kHz. We assume that the detection range required in V2X is 400 m and the highest speed of vehicle to be detected is 120 km/h.

In order to get satisfactory sidelobe in pulse compression, the distance corresponding to CP length should be greater than 400 m. Based on calculation, the length of CP is set as 32. When designing the ideal radar waveform, we minimize the ambiguity function of an interested delay-Doppler zone. In view of the sampling frequency, the Doppler frequency brought by the highest speed of 120 km/h can be neglected. Therefore, we set normalized Doppler frequency as 0 for the interested range-Doppler zone. Based on the sampling interval and detection range of 400 m, then the delay range is $[-32,32]$ for Nyquist-rate sampling and $[-64,64]$ for oversampling in the interested delay-Doppler zone. The constellation chosen for the communication users is quadrature phase shift keying (QPSK).

%In the numerical simulations, some basic simulation parameters are listed in Table~\ref{tab1}.
\begin{table}[!t]
\caption{Basic simulation parameters}
\label{tab1}
\centering
\begin{tabular}{ccc}
\hline
		Meaning & Symbol& Value \\
\hline
		Frequency of system	& ${{f}_{c}}$ &28GHz\\
		Subcarrier interval	& $\Delta f$& 300kHz \\
		Number of antennas & ${{N}_{t}}$& 8 \\
		Number of subcarriers &	${{N}_{s}}$	& 40 \\
		Length of CP &	${{N}_{cp}}$ &	32\\
		Number of users and targets	& $U$ & 2\\
		Angles of the targets & ${{\Omega }_{d}}$ & $\{-30{}^\circ ,30{}^\circ \}$	\\ 
		Oversampling rate & $\gamma $ & 2\\
		Number of channel taps & $T$ &4\\
		Rician factor of Rician channel & $K$ &	1\\
		Number of Monte Carlo simulations &	$N\_mc$ & 2000\\
\hline
\end{tabular}
\end{table}

\subsection{Convergence analysis with different initiation points in ADMM}
\label{s4-1}

\begin{figure}[!ht]
	\centering
	\subfloat[\label{fig1.a}objective function]{\includegraphics[width=.5\linewidth]{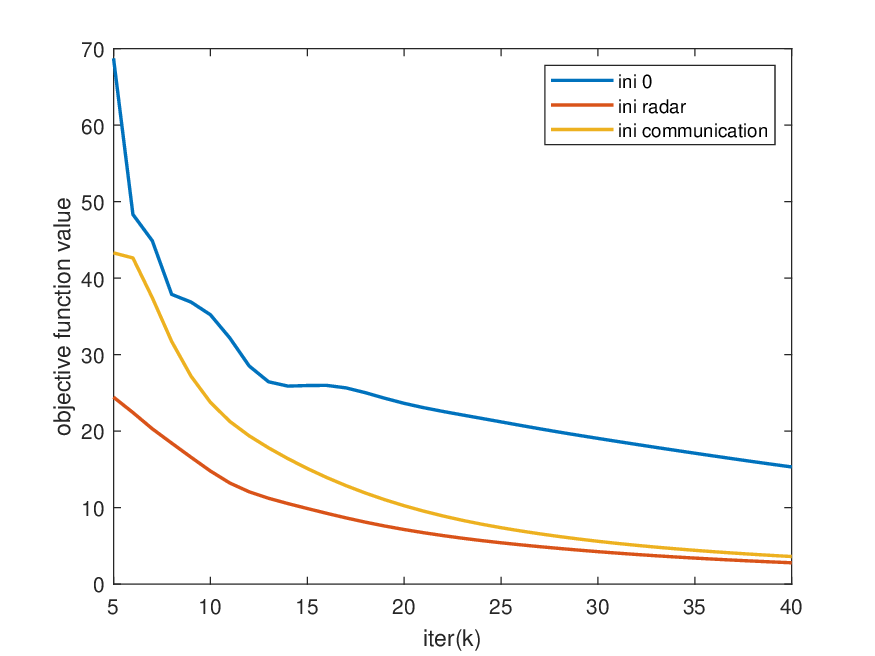}}%
	\subfloat[\label{fig1.b}Lagrange function]{\includegraphics[width=.5\linewidth]{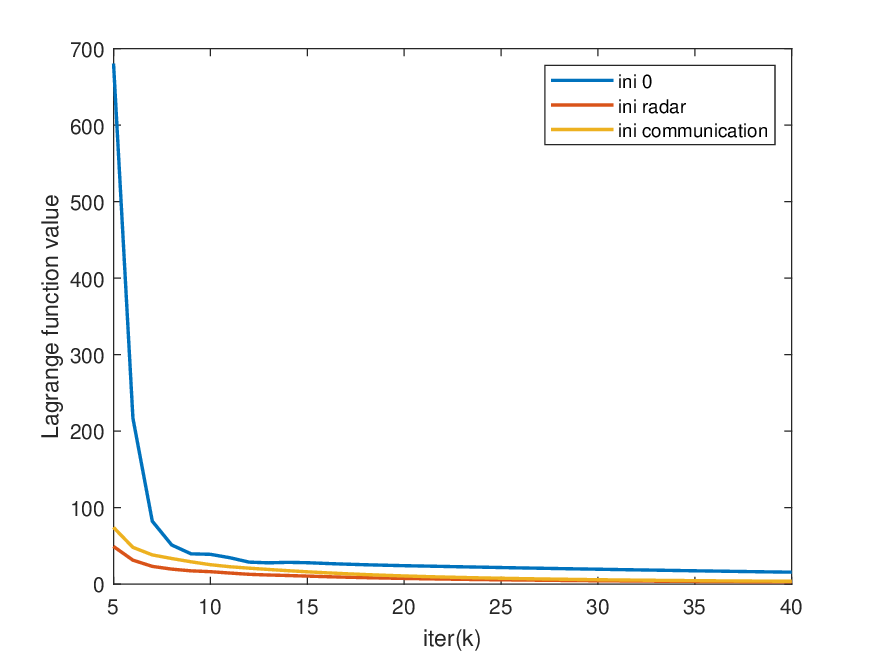}}
	\caption{Convergence of objective function and Lagrange function in ADMM with different initiation points}
	\label{fig1}
\end{figure}

\begin{figure}[!ht]
	\centering
	\subfloat[\label{fig2.a}communication term]{\includegraphics[width=.5\linewidth]{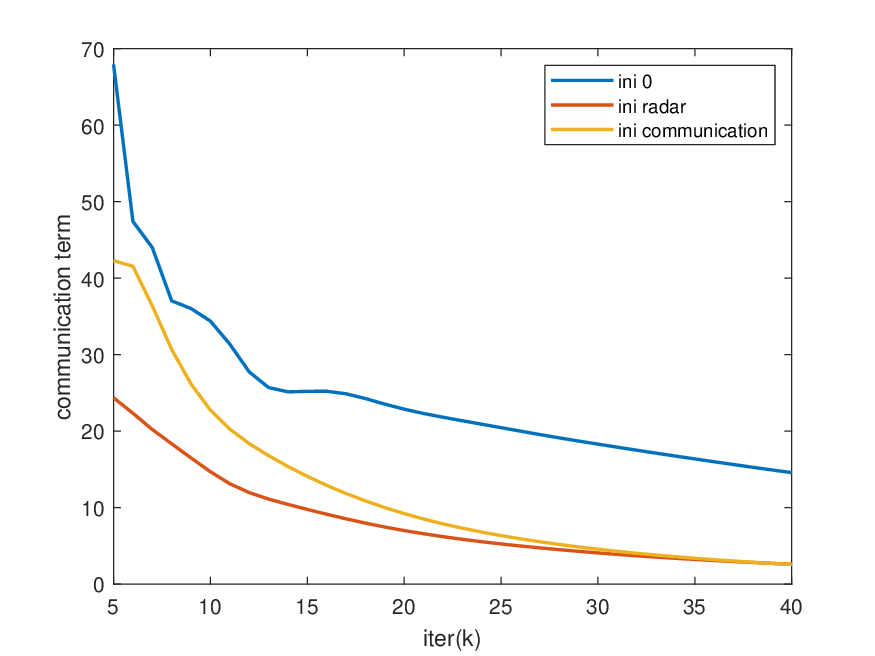}}%
	\subfloat[\label{fig2.b}radar term]{\includegraphics[width=.5\linewidth]{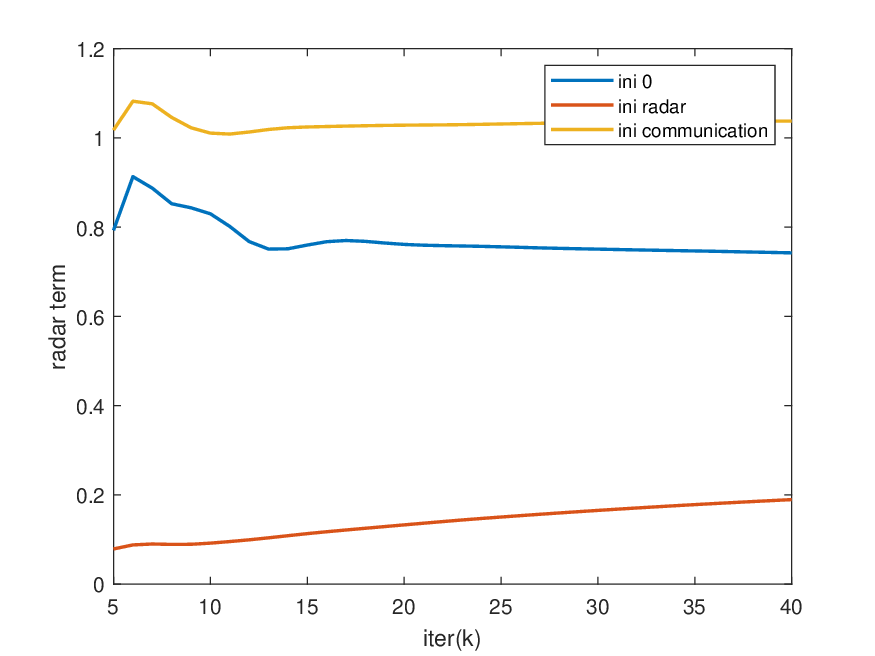}}
	\caption{Convergence of communication term and radar term in ADMM with different initiation points}
	\label{fig2}
\end{figure}

Figure \ref{fig1} and Figure \ref{fig2} show the convergence of ADMM with different initiation points. The `ini 0’ means the initiation point of ADMM is set to zero. The `ini radar’ represents that the initiation point of ADMM is set as the ideal radar waveform which is designed in the previous steps. The ‘ini communication’ shows that the initiation point of ADMM is set to the ideal communication waveform which is designed in an optimization only considering MUI. In terms of the objective function value, values corresponding to zero initiation point and radar initiation point converge at the same point. Meanwhile, the objective function value of communication initiation point converges to a higher value.

Figure \ref{fig2} gives a more detailed view of the convergence, through which we can see the convergence of communication term and radar term separately. It can be observed that communication terms corresponding to different initiation points all converge to the same point. As to the radar terms, the terms of zero initiation point and radar initiation point achieve the same value, while the term of communication initiation point converges to a higher value. Therefore, it is noteworthy that the initiation point of ADMM is the ideal radar waveform in the following simulations.

\subsection{Performance Comparison with different tradeoff between radar and communication}
\label{s4-2}

\begin{figure}[!ht]
	\centering
	\subfloat[\label{fig3.a}Nyquist Sampling]{\includegraphics[width=.5\linewidth,height=0.7\linewidth]{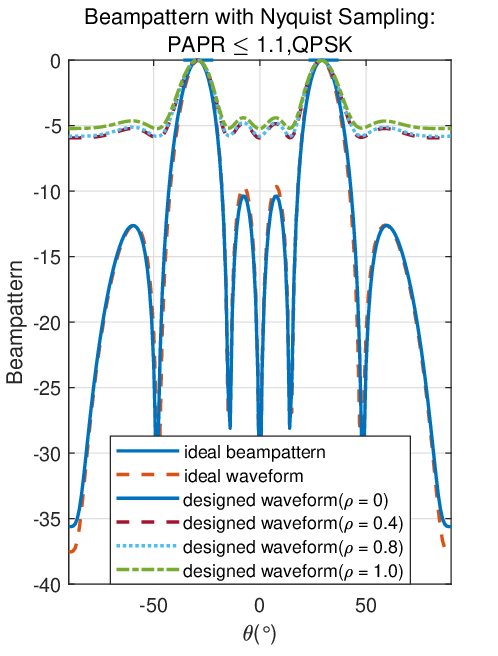}}
	\subfloat[\label{fig3.b}Oversampling]{\includegraphics[width=.5\linewidth,height=0.7\linewidth]{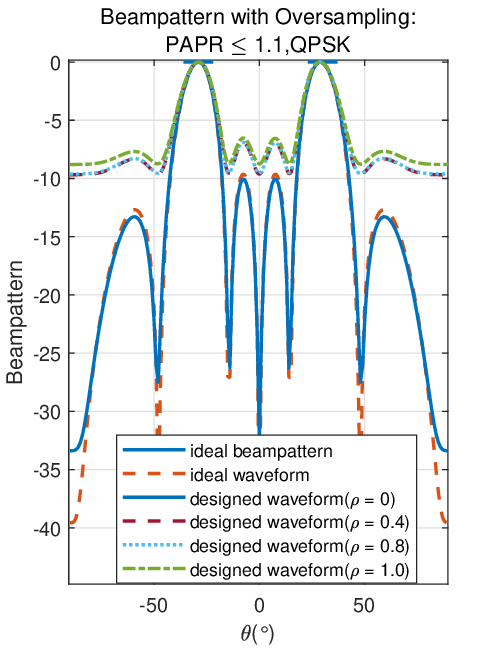}}
	\caption{Beam pattern of the waveform designed with different $\rho $}
	\label{fig3}
\end{figure}

\begin{figure}[!ht]
	\centering
	\subfloat[\label{fig4.a}Nyquist Sampling]{\includegraphics[width=.5\linewidth,height=0.7\linewidth]{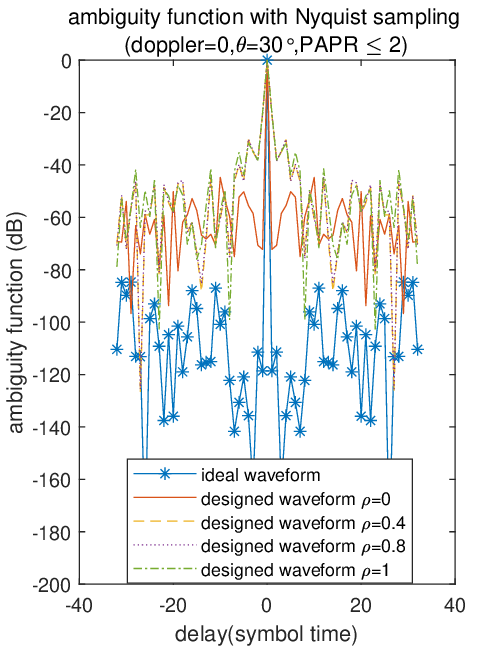}}
	\subfloat[\label{fig4.b}Oversampling]{\includegraphics[width=.5\linewidth,height=0.7\linewidth]{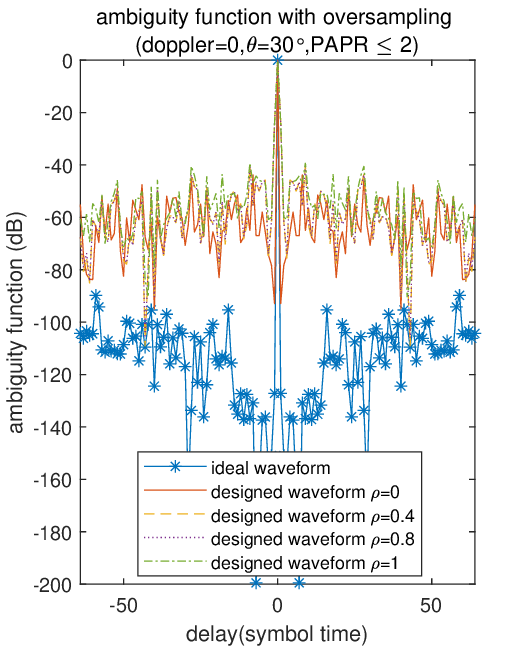}}
	\caption{Ambiguity function of the waveform designed with different $\rho $}
	\label{fig4}
\end{figure}

\begin{figure}[!ht]
	\centering
	\subfloat[\label{fig5.a}Nyquist Sampling]{\includegraphics[width=.5\linewidth,height=0.7\linewidth]{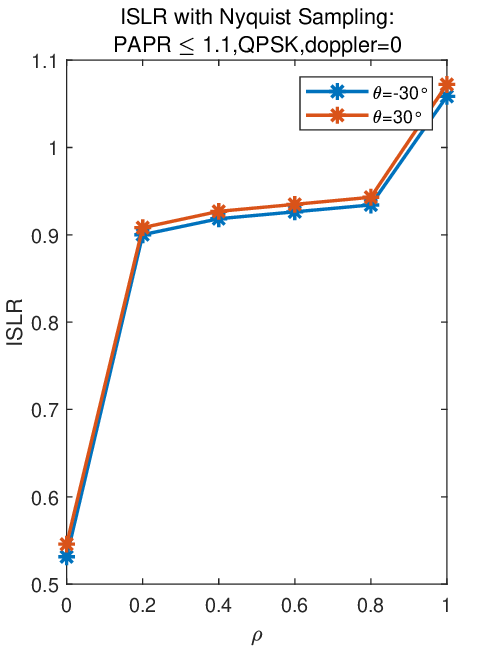}}
	\subfloat[\label{fig5.b}Oversampling]{\includegraphics[width=.5\linewidth,height=0.7\linewidth]{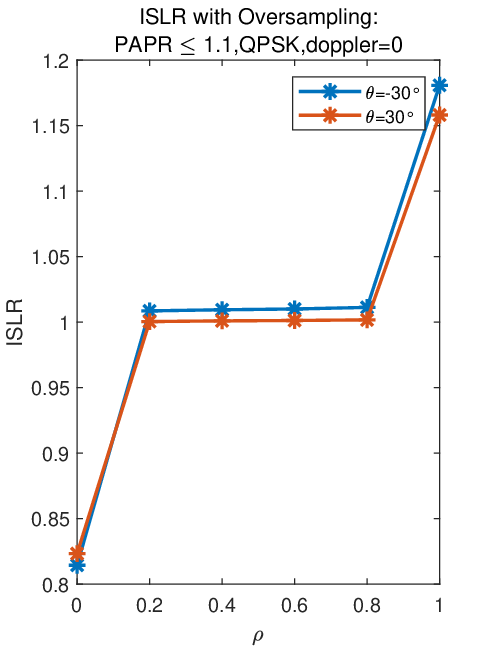}}
	\caption{ISLR of the waveform designed with different $\rho $}
	\label{fig5}
\end{figure}

\begin{figure}[!ht]
	\centering
	\subfloat[\label{fig6.a}Nyquist Sampling]{\includegraphics[width=.5\linewidth,height=0.7\linewidth]{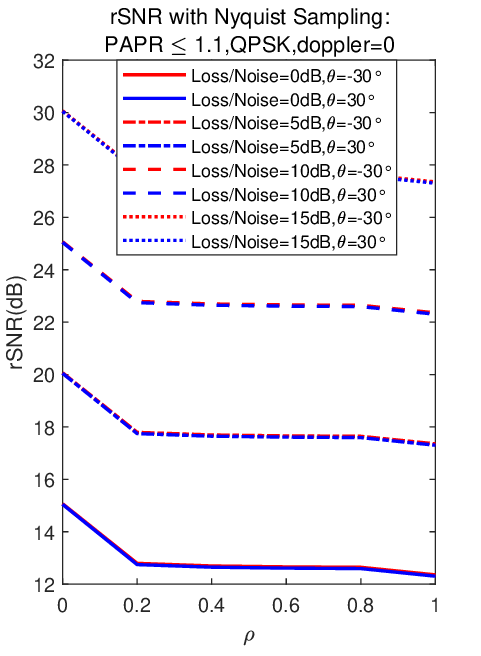}}
	\subfloat[\label{fig6.b}Oversampling]{\includegraphics[width=.5\linewidth,height=0.7\linewidth]{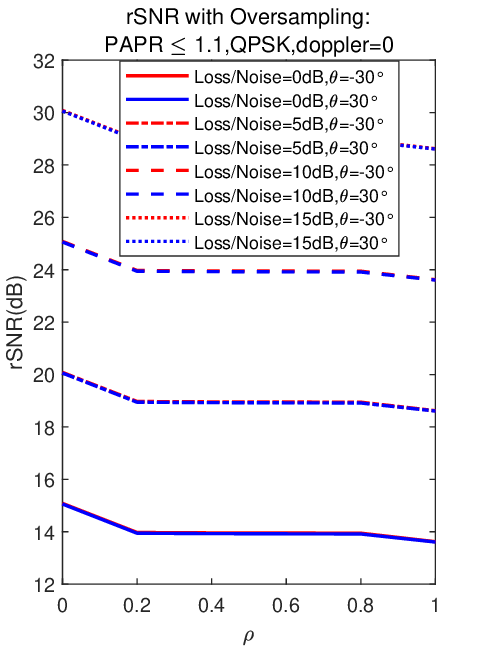}}
	\caption{rSNR of the waveform designed with different $\rho $}
	\label{fig6}
\end{figure}

\begin{figure}[!ht]
	\centering
	\subfloat[\label{fig7.a}Nyquist Sampling]{\includegraphics[width=.5\linewidth,height=0.7\linewidth]{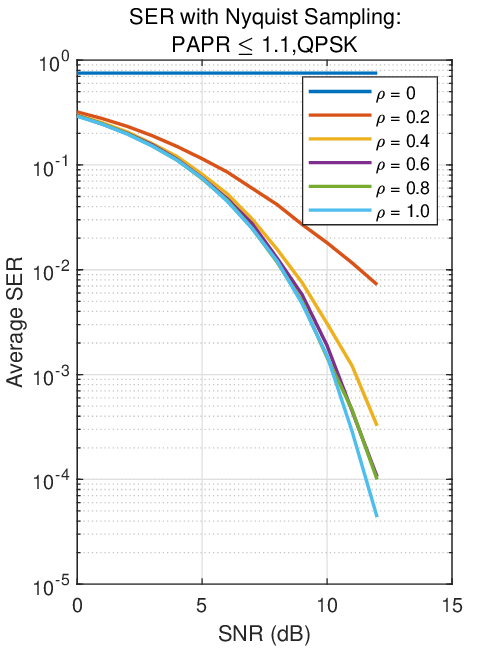}}
	\subfloat[\label{fig7.b}Oversampling]{\includegraphics[width=.5\linewidth,height=0.7\linewidth]{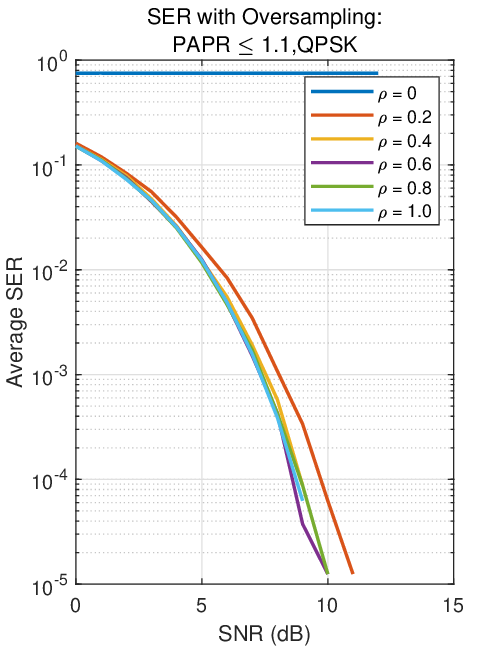}}
	\caption{Average SER of the waveform designed with different $\rho $}
	\label{fig7}
\end{figure}

\begin{figure}[!ht]
	\centering
	\subfloat[\label{fig8.a}Nyquist Sampling]{\includegraphics[width=.5\linewidth,height=0.7\linewidth]{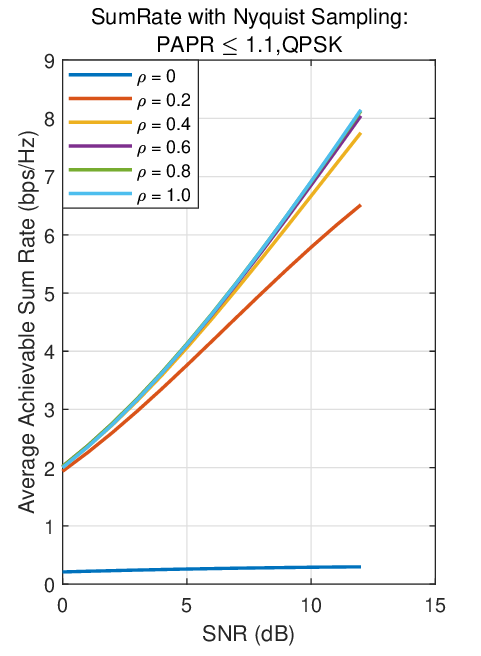}}
	\subfloat[\label{fig8.b}Oversampling]{\includegraphics[width=.5\linewidth,height=0.7\linewidth]{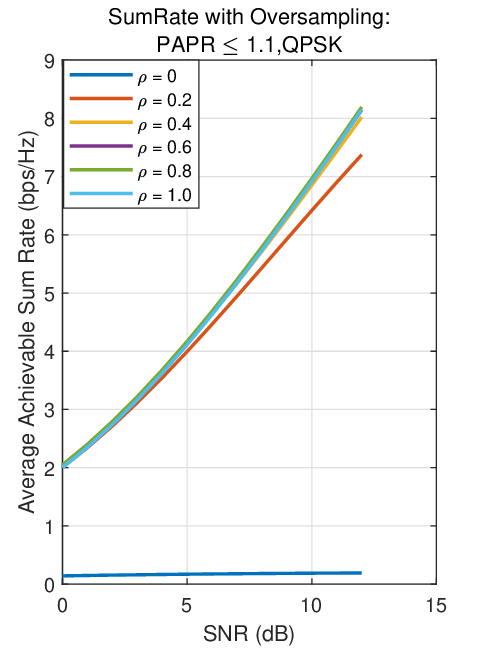}}
	\caption{Average achievable sum rate of the waveform designed with different $\rho $}
	\label{fig8}
\end{figure}

Figure \ref{fig3} presents the beam pattern of the waveforms designed with different tradeoff between radar and communication. It also contains the ideal beam pattern and the beam pattern of the designed ideal radar waveform. The beam pattern of the ideal radar waveform is closest to the ideal beam pattern, so is the beam pattern of designed waveform with $\rho =0$. It is easy to see that the gap between the beam pattern of the designed waveform and the ideal one increases with $\rho $. As can be seen in Figure \ref{fig4}, the sidelobe of the ambiguity function gets lower with a smaller $\rho $. Figure \ref{fig5} explicitly shows that ISLR increases with $\rho $. With $\rho $ increasing from 0 to 1, the design puts more priority to communication while imposing less restriction on the similarity of ideal radar waveform. As a result, the radar performance degrades with the increase of $\rho $. The signal to noise ratio (SNR) of the received echo decreased slightly with the increasing of $\rho $ in Figure \ref{fig6}. Figure \ref{fig6} also illustrates that the received SNR is proportional to the ratio of path loss to noise power. The `Loss/Noise' in Figure \ref{fig6} means the ratio of path loss to noise power. Because of the symmetry of the design to angles, the lines of  $-30{}^\circ $ and $30{}^\circ $ coincide to form a single line. We depict the curves of SER and average sum rate versus $\rho $ in Figure \ref{fig7} and Figure \ref{fig8}, respectively. As $\rho $ approaches to 1, the average SER decreases accordingly, meanwhile, the average sum rate increases. It is obvious that communication performance has higher priority in design when $\rho $ is close to 1 and radar performance dominates the design when $\rho $ approaches 0. 

In terms of Nyquist sampling and oversampling, the beam pattern is more close to the ideal beam pattern in the case of oversampling, which can be observed in Figure \ref{fig3}. According to Figure \ref{fig7} and Figure \ref{fig8}, the communication performance is also greater in the case of oversampling compared to Nyquist sampling.

\subsection{Performance Comparison with different PAPR constraints}
\label{s4-3}

\begin{figure}[!ht]
	\centering
	\includegraphics[width=0.4\textwidth]{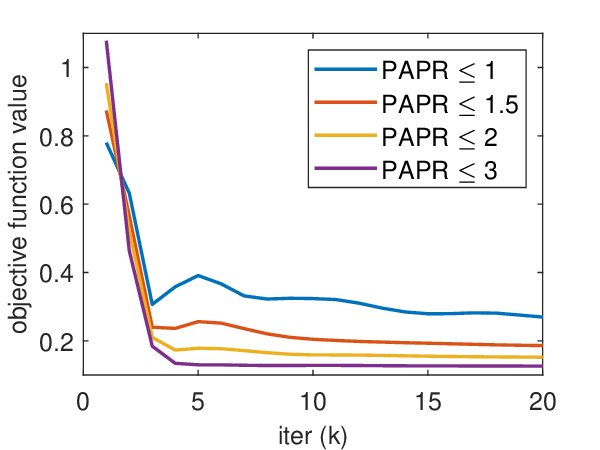}
	\caption{Convergence of objective function in ADMM with different PAPR constraints}
	\label{fig9}
\end{figure}

\begin{figure}[!ht]
	\centering
	\subfloat[\label{fig10.a}communication term]{\includegraphics[width=.5\linewidth]{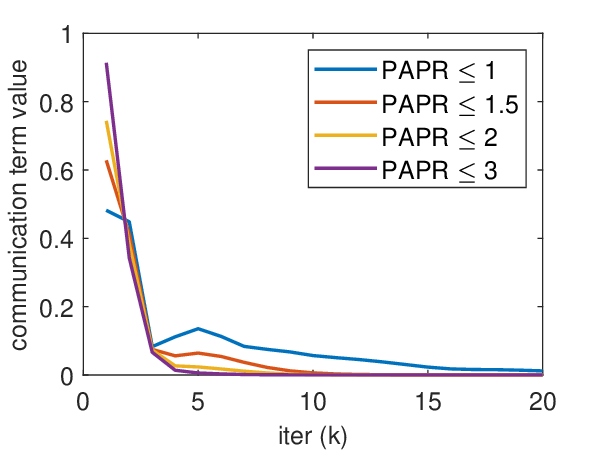}}%
	\subfloat[\label{fig10.b}radar term]{\includegraphics[width=.5\linewidth]{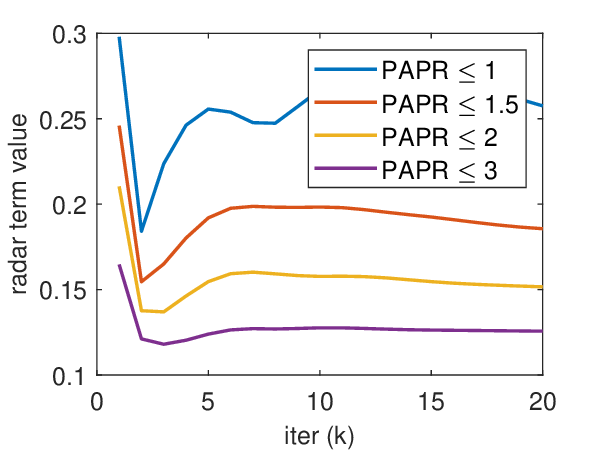}}
	\caption{Convergence of communication term and radar term in ADMM with different PAPR constraints}
	\label{fig10}
\end{figure}

Figure \ref{fig9} and Figure \ref{fig10} present the convergence of ADMM with different PAPR constraints. It is clear in Figure \ref{fig9} that the objective function converges to a smaller value with the relaxation of PAPR constraints. Figure \ref{fig10} gives a deeper insight into the convergence of ADMM by displaying the convergence curves of communication term and radar term, respectively. We can see that the communication term always converges to the same value with different PAPR constraints. On the contrary, the radar term converges to higher value with more strict PAPR constraints. The convergence curve of communication term can help explain why the average SER and average achievable sum rate remain almost unchanged when tightening PAPR constraints.
 
\begin{figure}[!ht]
	\centering
	\subfloat[\label{fig11.a}Nyquist Sampling]{\includegraphics[width=.5\linewidth,height=0.7\linewidth]{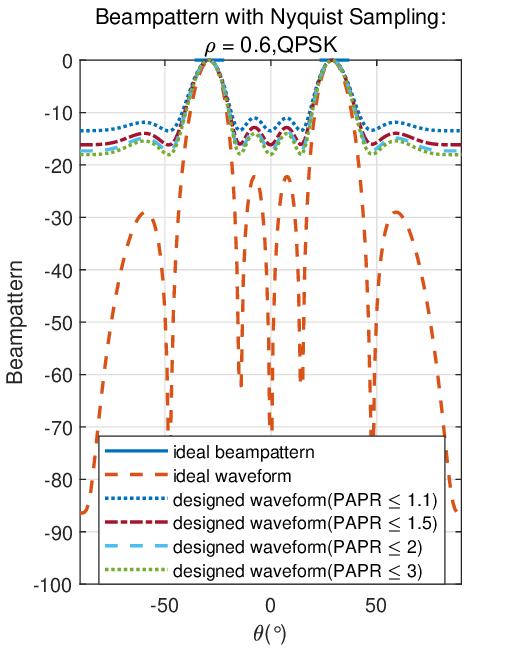}}
	\subfloat[\label{fig11.b}Oversampling]{\includegraphics[width=.5\linewidth,height=0.7\linewidth]{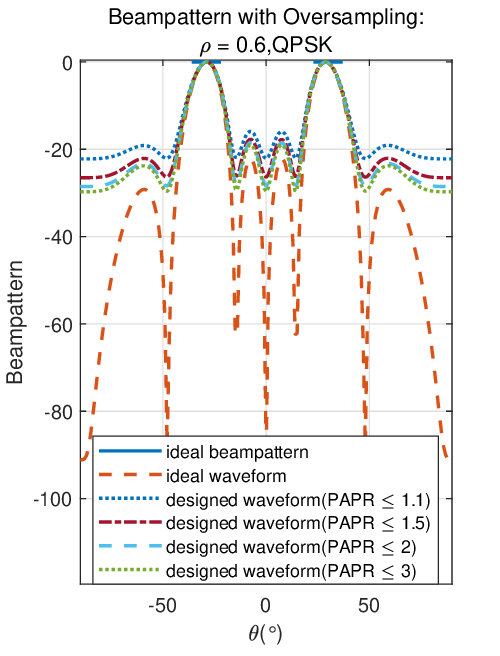}}
	\caption{Beam pattern of the waveform designed with different PAPR constraints}
	\label{fig11}
\end{figure}

\begin{figure}[!ht]
	\centering
	\subfloat[\label{fig12.a}Nyquist Sampling]{\includegraphics[width=.5\linewidth,height=0.7\linewidth]{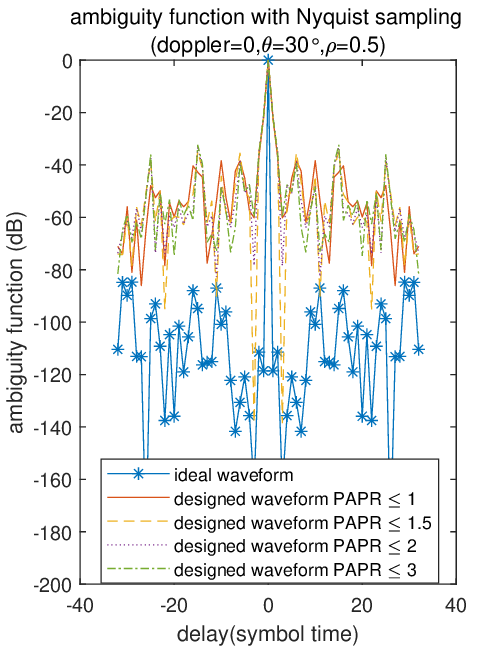}}
	\subfloat[\label{fig12.b}Oversampling]{\includegraphics[width=.5\linewidth,height=0.7\linewidth]{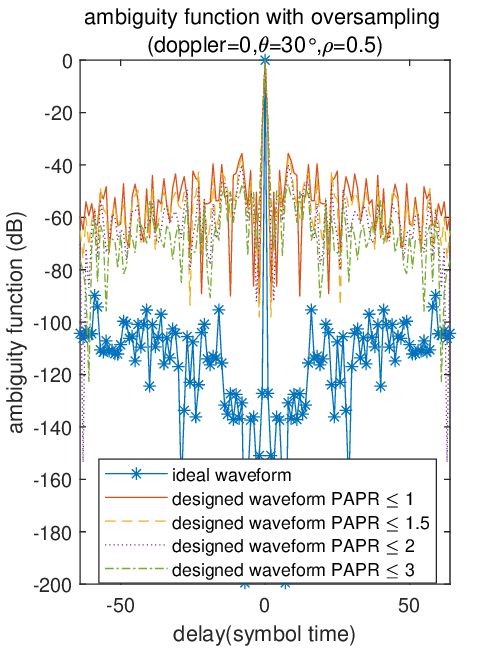}}
	\caption{Ambiguity function of the waveform designed with different PAPR constraints}
	\label{fig12}
\end{figure}

\begin{figure}[!ht]
	\centering
	\subfloat[\label{fig13.a}Nyquist Sampling]{\includegraphics[width=.5\linewidth,height=0.7\linewidth]{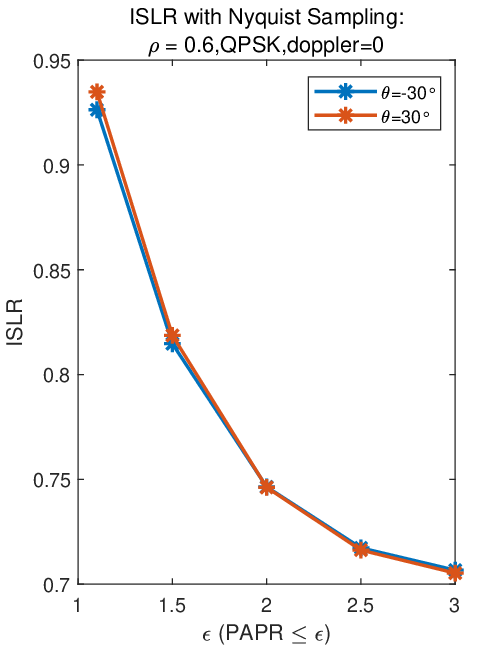}}
	\subfloat[\label{fig13.b}Oversampling]{\includegraphics[width=.5\linewidth,height=0.7\linewidth]{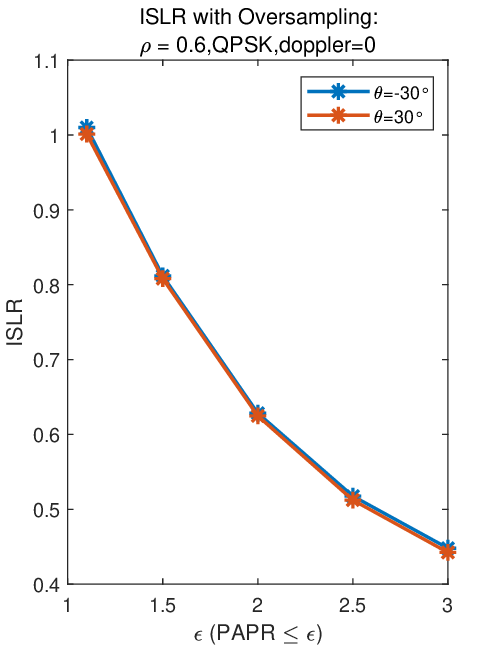}}
	\caption{ISLR of the waveform designed with different PAPR constraints}
	\label{fig13}
\end{figure}

\begin{figure}[!ht]
	\centering
	\subfloat[\label{fig14.a}Nyquist Sampling]{\includegraphics[width=.5\linewidth,height=0.7\linewidth]{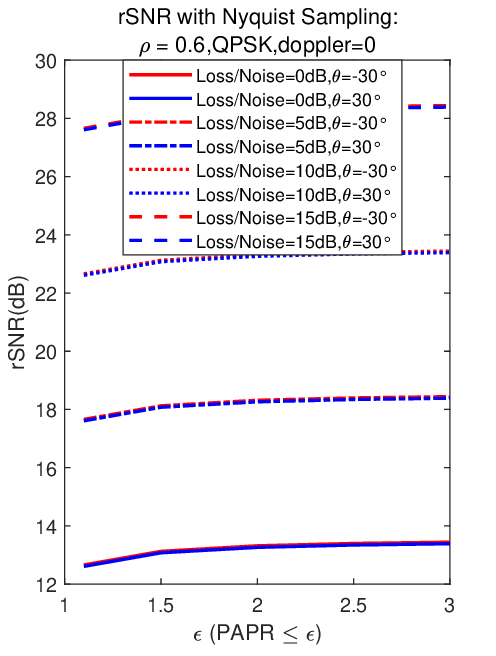}}
	\subfloat[\label{fig14.b}Oversampling]{\includegraphics[width=.5\linewidth,height=0.7\linewidth]{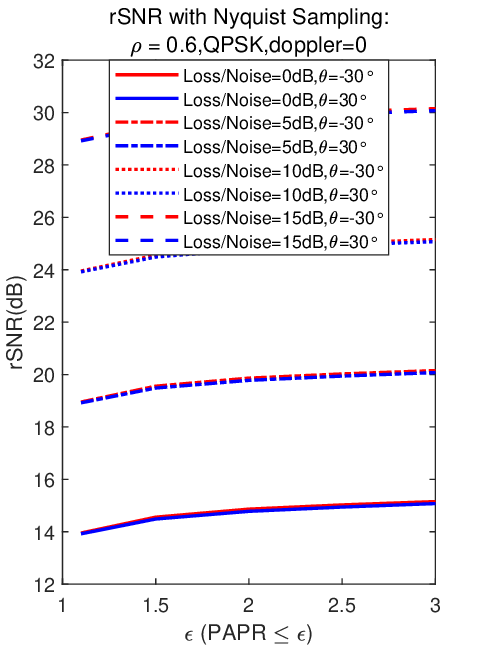}}
	\caption{rSNR of the waveform designed with different PAPR constraints}
	\label{fig14}
\end{figure}

\begin{figure}[!ht]
	\centering
	\subfloat[\label{fig15.a}Nyquist Sampling]{\includegraphics[width=.5\linewidth,height=0.7\linewidth]{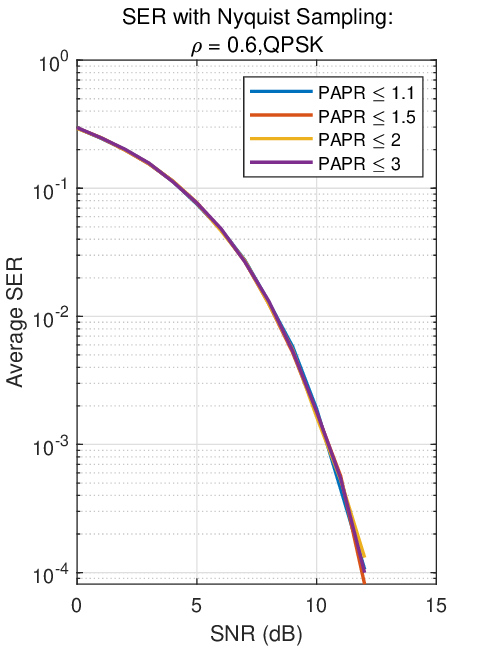}}
	\subfloat[\label{fig15.b}Oversampling]{\includegraphics[width=.5\linewidth,height=0.7\linewidth]{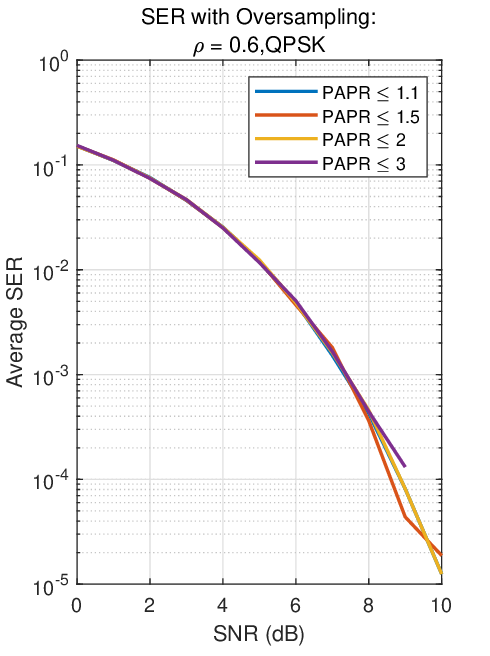}}
	\caption{Average SER of the waveform designed with different PAPR constraints}
	\label{fig15}
\end{figure}

\begin{figure}[!ht]
	\centering
	\subfloat[\label{fig16.a}Nyquist Sampling]{\includegraphics[width=.5\linewidth,height=0.7\linewidth]{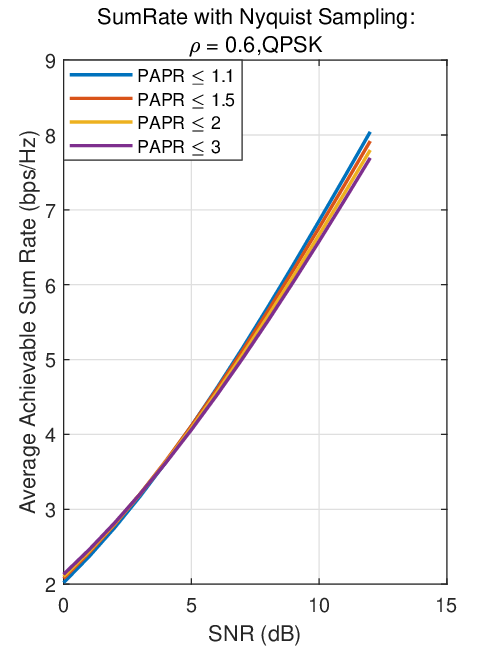}}
	\subfloat[\label{fig16.b}Oversampling]{\includegraphics[width=.5\linewidth,height=0.7\linewidth]{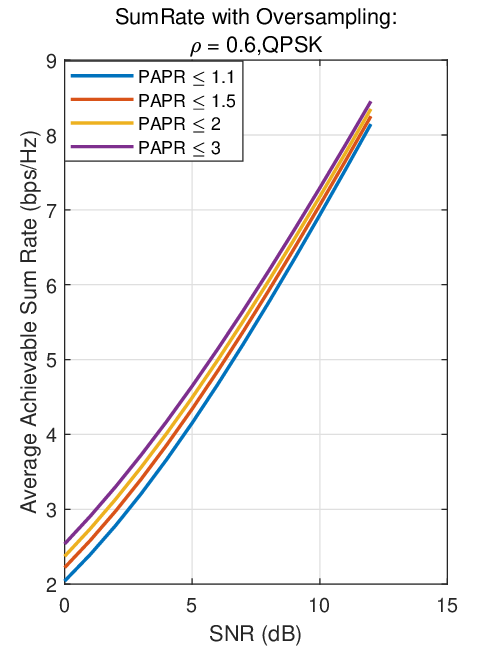}}
	\caption{Average achievable sum rate of the waveform designed with different PAPR constraints}
	\label{fig16}
\end{figure}

It is clearly shown in Figure \ref{fig11} to Figure \ref{fig14} that better radar performance can be achieved with more relaxed PAPR constraints. However, the deterioration of communication performance, which is reflected in Figure \ref{fig15} and Figure \ref{fig16}, is subtle with tighter PAPR constraints.
It proves that our algorithm can achieve satisfactory communication performance even with the PAPR constraint tightened to a certain extent. It can be explained by Figure \ref{fig10}, where the communication term of the objective function always converges to the same value with different PAPR constraints. The convergence curve of communication term in ADMM helps explain why the average SER and average achievable sum rate remain almost unchanged when tightening PAPR constraints.
 Besides, it is easy to notice in Figure \ref{fig16} that the gaps between the curves corresponding to different PAPR constraints are wider in the case with oversampling. The reason lies in that PAPR is measured more accurately with oversampling. Compared to Nyquist sampling, the system performance has more obvious deterioration when tightening the PAPR constraints with sufficient sampling. 

\section{conclusion}
\label{CONCLUSION}

In this paper, we formulated a weighted optimization problem to balance the performance of radar and communication systems. Through minimizing MUI in the objective function, we improved the communication performance. By introducing similarity between ISAC waveform and ideal radar waveform to the objective function, we guaranteed the radar performance in terms of beam pattern and ambiguity function. Considering the limitation of power amplifier, we set the PAPR of each antenna below a threshold by adding constraints to the optimization problem. In the design of the ideal radar waveform, we employed L-BFGS which enabled the algorithm to obtain the ideal waveform quickly with limited memory space. For the design of the ISAC waveform, we solved the optimization problem with ADMM. The iteration of each variable has a closed-form solution, which guaranteed fast processing. Moreover, we conducted numerical simulations. The results validated that the proposed method can improve the performance of the ISAC waveform in both radar sensing and communication.

\appendix

\subsection{Proof of proposition 1}
\label{app:a}
The  $\left( \text{m}+1 \right)$-th update of $\mathbf{s}$ gives
       \begin{equation}
\begin{aligned}
  & -\eta {{\mathbf{T}}^{T}}\left( {{\mathbf{\xi }}^{\left( m+1 \right)}}-\mathbf{T}{{\mathbf{s}}^{\left( m+1 \right)}}+{{{\mathbf{\tilde{\lambda }}}}^{\left( m \right)}} \right) \\ 
 & +\nabla f\left( {{\mathbf{s}}^{\left( m+1 \right)}} \right)=0 \\ 
\end{aligned}
       \end{equation}
       
       According to ${{\mathbf{\tilde{\lambda }}}^{\left( m+1 \right)}}={{\mathbf{\tilde{\lambda }}}^{\left( m \right)}}+{{\mathbf{\xi }}^{\left( m+1 \right)}}-\mathbf{T}{{\mathbf{s}}^{\left( m+1 \right)}}$, the equation can be expressed as
       \begin{equation}
       	\nabla f\left( {{\mathbf{s}}^{\left( m+1 \right)}} \right)=\eta {{\mathbf{T}}^{\text{T}}}{{\mathbf{\tilde{\lambda }}}^{\left( m+1 \right)}}.
       \end{equation}
       It can be easily proved that $\nabla f$ is Lipschitz continuous with parameter ${{L}_{f}}=\left| {{\lambda }_{\max }}\left( {{\mathbf{Q}}^{T}}\mathbf{Q} \right) \right|$. Based on the definition of Lipschitz continuity, we have
        \begin{equation}
\begin{aligned}
  & {{\left\| \nabla f\left( {{\mathbf{s}}^{\left( m+1 \right)}} \right)-\nabla f\left( {{\mathbf{s}}^{\left( m \right)}} \right) \right\|}_{2}} \\ 
 & \le {{L}_{f}}{{\left\| {{\mathbf{s}}^{\left( m+1 \right)}}-{{\mathbf{s}}^{\left( m \right)}} \right\|}_{2}} \\ 
\end{aligned}
       \end{equation}

       According to $\nabla f\left( {{\mathbf{s}}^{\left( m+1 \right)}} \right)=\eta {{\mathbf{T}}^{\text{T}}}{{\mathbf{\tilde{\lambda }}}^{\left( m+1 \right)}}$, the inequality above can be transformed to
       \begin{equation}
\begin{aligned}
  & {{\eta }^{2}}\left\| {{\mathbf{T}}^{T}}\left( {{{\mathbf{\tilde{\lambda }}}}^{\left( m+1 \right)}}-{{{\mathbf{\tilde{\lambda }}}}^{\left( m \right)}} \right) \right\|_{2}^{2} \\ 
 & \le L_{f}^{2}\left\| {{\mathbf{s}}^{\left( m+1 \right)}}-{{\mathbf{s}}^{\left( m \right)}} \right\|_{2}^{2} \\ 
\end{aligned}
       \end{equation}
       As long as $\varepsilon $ is small enough and $\alpha $ is large enough, we can deduce the following inequality as
       \begin{equation}
    \begin{aligned}
    & {{\eta }^{2}}\left\| {{\mathbf{T}}^{T}}\left( {{{\mathbf{\tilde{\lambda }}}}^{\left( m+1 \right)}}-{{{\mathbf{\tilde{\lambda }}}}^{\left( m \right)}} \right) \right\|_{2}^{2}\\
    &+{{\eta }^{2}}{{\varepsilon }^{2}}\left\| {{{\mathbf{\tilde{\lambda }}}}^{\left( m+1 \right)}}-{{{\mathbf{\tilde{\lambda }}}}^{\left( m \right)}} \right\|_{2}^{2}\\ 
    & \le {{\alpha }^{2}}L_{f}^{2}\left\| {{\mathbf{s}}^{\left( m+1 \right)}}-{{\mathbf{s}}^{\left( m \right)}} \right\|_{2}^{2} \\ 
    \end{aligned}
       \end{equation}
       The first term of the left satisfies
       \begin{equation}
       	\begin{aligned}
       		& {{\eta }^{2}}\left\| {{\mathbf{T}}^{\text{T}}}\left( {{{\mathbf{\tilde{\lambda }}}}^{\left( m+1 \right)}}-{{{\mathbf{\tilde{\lambda }}}}^{\left( m \right)}} \right) \right\|_{2}^{2} \\ 
       		& ={{\eta }^{2}}{{\left( {{{\mathbf{\tilde{\lambda }}}}^{\left( m+1 \right)}}-{{{\mathbf{\tilde{\lambda }}}}^{\left( m \right)}} \right)}^{H}}\mathbf{T}{{\mathbf{T}}^{\text{T}}}\left( {{{\mathbf{\tilde{\lambda }}}}^{\left( m+1 \right)}}-{{{\mathbf{\tilde{\lambda }}}}^{\left( m \right)}} \right) \\ 
       		& \ge {{\eta }^{2}}{{\lambda }_{\min }}\left( \mathbf{T}{{\mathbf{T}}^{\text{T}}} \right)\left\| {{{\mathbf{\tilde{\lambda }}}}^{\left( m+1 \right)}}-{{{\mathbf{\tilde{\lambda }}}}^{\left( m \right)}} \right\|_{2}^{2} . 
       	\end{aligned}
       \end{equation}
       Because ${{\lambda }_{\min }}\left( \mathbf{T}{{\mathbf{T}}^{T}} \right)=0$, we have the inequality as follows
       \begin{equation}
\begin{aligned}
  & \left\| {{{\mathbf{\tilde{\lambda }}}}^{\left( m+1 \right)}}-{{{\mathbf{\tilde{\lambda }}}}^{\left( m \right)}} \right\|_{2}^{2} \\ 
 & \le \frac{{{\alpha }^{2}}L_{f}^{2}}{{{\eta }^{2}}{{\varepsilon }^{2}}}\left\| {{\mathbf{s}}^{\left( m+1 \right)}}-{{\mathbf{s}}^{\left( m \right)}} \right\|_{2}^{2} \\ 
\end{aligned}
       \end{equation}

\subsection{Proof of proposition 2}
\label{app:b}
The successive difference of the augmented Lagrange function can be split by
       \begin{equation}
       	\begin{aligned}
       		& {{L}_{\eta }}\left( {{\mathbf{s}}^{\left( m+1 \right)}},{{\mathbf{\xi }}^{\left( m+1 \right)}}\text{,}{{{\mathbf{\tilde{\lambda }}}}^{\left( m+1 \right)}} \right)-{{L}_{\eta }}\left( {{\mathbf{s}}^{\left( m \right)}},{{\mathbf{\xi }}^{\left( m \right)}}\text{,}{{{\mathbf{\tilde{\lambda }}}}^{\left( m \right)}} \right)= \\ 
       		& {{L}_{\eta }}\left( {{\mathbf{s}}^{\left( m+1 \right)}},{{\mathbf{\xi }}^{\left( m+1 \right)}}\text{,}{{{\mathbf{\tilde{\lambda }}}}^{\left( m+1 \right)}} \right)-{{L}_{\eta }}\left( {{\mathbf{s}}^{\left( m+1 \right)}},{{\mathbf{\xi }}^{\left( m+1 \right)}}\text{,}{{{\mathbf{\tilde{\lambda }}}}^{\left( m \right)}} \right) \\ 
       		& +{{L}_{\eta }}\left( {{\mathbf{s}}^{\left( m+1 \right)}},{{\mathbf{\xi }}^{\left( m+1 \right)}}\text{,}{{{\mathbf{\tilde{\lambda }}}}^{\left( m \right)}} \right)-{{L}_{\eta }}\left( {{\mathbf{s}}^{\left( m \right)}},{{\mathbf{\xi }}^{\left( m+1 \right)}}\text{,}{{{\mathbf{\tilde{\lambda }}}}^{\left( m \right)}} \right) \\ 
       		& +{{L}_{\eta }}\left( {{\mathbf{s}}^{\left( m \right)}},{{\mathbf{\xi }}^{\left( m+1 \right)}}\text{,}{{{\mathbf{\tilde{\lambda }}}}^{\left( m \right)}} \right)-{{L}_{\eta }}\left( {{\mathbf{s}}^{\left( m \right)}},{{\mathbf{\xi }}^{\left( m \right)}}\text{,}{{{\mathbf{\tilde{\lambda }}}}^{\left( m \right)}} \right). \\ 
       	\end{aligned}
       \end{equation}
       
       The first term is bounded by
       \begin{equation}
       	\begin{aligned}
       		& {{L}_{\eta }}\left( {{\mathbf{s}}^{\left( m+1 \right)}},{{\mathbf{\xi }}^{\left( m+1 \right)}}\text{,}{{{\mathbf{\tilde{\lambda }}}}^{\left( m+1 \right)}} \right)-{{L}_{\eta }}\left( {{\mathbf{s}}^{\left( m+1 \right)}},{{\mathbf{\xi }}^{\left( m+1 \right)}}\text{,}{{{\mathbf{\tilde{\lambda }}}}^{\left( m \right)}} \right) \\ 
       		& =\eta \left[ {{\left( {{{\mathbf{\tilde{\lambda }}}}^{\left( m+1 \right)}}-{{{\mathbf{\tilde{\lambda }}}}^{\left( m \right)}} \right)}^{T}}\left( {{\mathbf{\xi }}^{\left( m+1 \right)}}-\mathbf{T}{{\mathbf{s}}^{\left( m+1 \right)}} \right) \right] \\ 
       		& =\eta {{\left( {{{\mathbf{\tilde{\lambda }}}}^{\left( m+1 \right)}}-{{{\mathbf{\tilde{\lambda }}}}^{\left( m \right)}} \right)}^{T}}\left( {{{\mathbf{\tilde{\lambda }}}}^{\left( m+1 \right)}}-{{{\mathbf{\tilde{\lambda }}}}^{\left( m \right)}} \right) \\ 
       		& =\eta \left\| {{{\mathbf{\tilde{\lambda }}}}^{\left( m+1 \right)}}-{{{\mathbf{\tilde{\lambda }}}}^{\left( m \right)}} \right\|_{2}^{2} \\ 
       		& \le \frac{{{\alpha }^{2}}L_{f}^{2}}{\eta {{\varepsilon }^{2}}}\left\| {{\mathbf{s}}^{\left( m+1 \right)}}-{{\mathbf{s}}^{\left( m \right)}} \right\|_{2}^{2} .\\ 
       	\end{aligned}
       \end{equation}
       
       We define the function $g\left( \mathbf{s},\mathbf{\xi },\mathbf{\tilde{\lambda }} \right)=f\left( \mathbf{s} \right)+\frac{\eta }{2}\left\| \mathbf{\xi }-\mathbf{Ts}+\mathbf{\tilde{\lambda }} \right\|_{2}^{2}$. $g\left( \mathbf{s},\mathbf{\xi },\mathbf{\tilde{\lambda }} \right)$ is convex with respect to $\mathbf{s}$ because
       \begin{equation}
       	\begin{aligned}
       		& {{\left[ \nabla g\left( {{\mathbf{s}}_{1}},\mathbf{\xi },\mathbf{\tilde{\lambda }} \right)-\nabla g\left( {{\mathbf{s}}_{2}},\mathbf{\xi },\mathbf{\tilde{\lambda }} \right) \right]}^{\text{T}}}\left( {{\mathbf{s}}_{1}}-{{\mathbf{s}}_{2}} \right) \\ 
       		& ={{\left( {{\mathbf{s}}_{1}}-{{\mathbf{s}}_{2}} \right)}^{\text{T}}}\left[ 2{{\mathbf{Q}}^{\text{T}}}\mathbf{Q}+\eta {{\mathbf{T}}^{\text{T}}}\mathbf{T} \right]\left( {{\mathbf{s}}_{1}}-{{\mathbf{s}}_{2}} \right) \\ 
       		& \ge \left[ 2{{\lambda }_{\min }}\left( {{\mathbf{Q}}^{\text{T}}}\mathbf{Q} \right)+\eta {{\lambda }_{\min }}\left( {{\mathbf{T}}^{\text{T}}}\mathbf{T} \right) \right]\left\| {{\mathbf{s}}_{1}}-{{\mathbf{s}}_{2}} \right\|_{2}^{2} \\ 
       		& =2{{\lambda }_{\min }}\left( {{\mathbf{Q}}^{\text{T}}}\mathbf{Q} \right)\left\| {{\mathbf{s}}_{1}}-{{\mathbf{s}}_{2}} \right\|_{2}^{2}. \\ 
       	\end{aligned}
       \end{equation}
       
       Meanwhile, we have
       \begin{equation}
       	\begin{aligned}
       		& \nabla g\left( {{\mathbf{s}}^{\left( m+1 \right)}},{{\mathbf{\xi }}^{\left( m+1 \right)}}\text{,}{{{\mathbf{\tilde{\lambda }}}}^{\left( m \right)}} \right) \\ 
       		& =\nabla f\left( {{\mathbf{s}}^{\left( m+1 \right)}} \right)\\
         &-\eta {{\mathbf{T}}^{\text{T}}}\left( {{\mathbf{\xi }}^{\left( m+1 \right)}}-\mathbf{T}{{\mathbf{s}}^{\left( m+1 \right)}}+{{{\mathbf{\tilde{\lambda }}}}^{\left( m \right)}} \right) \\ 
       		& =\nabla f\left( {{\mathbf{s}}^{\left( m+1 \right)}} \right)-\eta {{\mathbf{T}}^{\text{T}}}{{{\mathbf{\tilde{\lambda }}}}^{\left( m+1 \right)}} \\ 
       		& =0 . 
       	\end{aligned}
       \end{equation}
       
       With the derivation above the second term is bounded by
       \begin{equation}
       	\begin{aligned}
       		& {{L}_{\eta }}\left( {{\mathbf{s}}^{\left( m+1 \right)}},{{\mathbf{\xi }}^{\left( m+1 \right)}}\text{,}{{{\mathbf{\tilde{\lambda }}}}^{\left( m \right)}} \right)-{{L}_{\eta }}\left( {{\mathbf{s}}^{\left( m \right)}},{{\mathbf{\xi }}^{\left( m+1 \right)}}\text{,}{{{\mathbf{\tilde{\lambda }}}}^{\left( m \right)}} \right) \\ 
       		& =g\left( {{\mathbf{s}}^{\left( m+1 \right)}},{{\mathbf{\xi }}^{\left( m+1 \right)}}\text{,}{{{\mathbf{\tilde{\lambda }}}}^{\left( m \right)}} \right)-g\left( {{\mathbf{s}}^{\left( m \right)}},{{\mathbf{\xi }}^{\left( m+1 \right)}}\text{,}{{{\mathbf{\tilde{\lambda }}}}^{\left( m \right)}} \right) \\ 
       		& \le {{\left[ \nabla g\left( {{\mathbf{s}}^{\left( m+1 \right)}},{{\mathbf{\xi }}^{\left( m+1 \right)}}\text{,}{{{\mathbf{\tilde{\lambda }}}}^{\left( m \right)}} \right) \right]}^{\text{T}}}\left( {{\mathbf{s}}^{\left( m+1 \right)}}-{{\mathbf{s}}^{\left( m \right)}} \right) \\ 
       		& -\frac{1}{2}\left[ 2{{\lambda }_{\min }}\left( {{\mathbf{Q}}^{\text{T}}}\mathbf{Q} \right)+\eta {{\lambda }_{\min }}\left( {{\mathbf{T}}^{\text{T}}}\mathbf{T} \right) \right]\left\| {{\mathbf{s}}^{\left( m+1 \right)}}-{{\mathbf{s}}^{\left( m \right)}} \right\|_{2}^{2} \\ 
       		& =-{{\lambda }_{\min }}\left( {{\mathbf{Q}}^{\text{T}}}\mathbf{Q} \right)\left\| {{\mathbf{s}}^{\left( m+1 \right)}}-{{\mathbf{s}}^{\left( m \right)}} \right\|_{2}^{2}.  
       	\end{aligned}
       \end{equation}
       
       Because of $\mathbf{\xi }=\arg \min {{L}_{\eta }}\left( {{\mathbf{s}}^{\left( m \right)}},\mathbf{\xi },{{{\mathbf{\tilde{\lambda }}}}^{\left( m \right)}} \right)$, the third term has
       \begin{equation}
\begin{aligned}
  & {{L}_{\eta }}\left( {{\mathbf{s}}^{\left( m \right)}},{{\mathbf{\xi }}^{\left( m+1 \right)}}\text{,}{{{\mathbf{\tilde{\lambda }}}}^{\left( m \right)}} \right) \\ 
 & -{{L}_{\eta }}\left( {{\mathbf{s}}^{\left( m \right)}},{{\mathbf{\xi }}^{\left( m \right)}}\text{,}{{{\mathbf{\tilde{\lambda }}}}^{\left( m \right)}} \right)\le 0 \\ 
\end{aligned}
       \end{equation}
       
       Based on the analysis above, we can obtain
       \begin{equation}
       	\begin{aligned}
       		& {{L}_{\eta }}\left( {{\mathbf{s}}^{\left( m+1 \right)}},{{\mathbf{\xi }}^{\left( m+1 \right)}}\text{,}{{{\mathbf{\tilde{\lambda }}}}^{\left( m+1 \right)}} \right)-{{L}_{\eta }}\left( {{\mathbf{s}}^{\left( m \right)}},{{\mathbf{\xi }}^{\left( m \right)}}\text{,}{{{\mathbf{\tilde{\lambda }}}}^{\left( m \right)}} \right) \\ 
       		& \le \left( \frac{{{\alpha }^{2}}L_{f}^{2}}{\eta {{\varepsilon }^{2}}}-{{\lambda }_{\min }}\left( {{\mathbf{Q}}^{\text{T}}}\mathbf{Q} \right) \right)\left\| {{\mathbf{s}}^{\left( m+1 \right)}}-{{\mathbf{s}}^{\left( m \right)}} \right\|_{2}^{2} . 
       	\end{aligned}
       	\label{eq21}
       \end{equation}
       
       If $\eta \ge \frac{{{\alpha }^{2}}L_{f}^{2}}{{{\varepsilon }^{2}}{{\lambda }_{\min }}\left( {{\mathbf{Q}}^{\text{T}}}\mathbf{Q} \right)}$, ${{L}_{\eta }}\left( {{\mathbf{s}}^{\left( m \right)}},{{\mathbf{\xi }}^{\left( m \right)}}\text{,}{{{\mathbf{\tilde{\lambda }}}}^{\left( m \right)}} \right)$ is a non-increasing sequence.

\subsection{Proof of proposition 3}
\label{app:c}
The Lagrange function can be expressed as 
       \begin{equation}
    \begin{aligned}
      & {{L}_{\eta }}\left( {{\mathbf{s}}^{\left( m \right)}},{{\mathbf{\xi }}^{\left( m \right)}}\text{,}{{{\mathbf{\tilde{\lambda }}}}^{\left( m \right)}} \right) \\ 
     & =\left\| \mathbf{Q}{{\mathbf{s}}^{\left( m \right)}}-\mathbf{\beta } \right\|_{2}^{2}+\frac{\eta }{2}{{\left\| {{\mathbf{\xi }}^{\left( m \right)}}-\mathbf{T}{{\mathbf{s}}^{\left( m \right)}} \right\|}_{2}^{2}} \\ 
     & +\frac{\eta }{2}\left( {{\mathbf{\xi }}^{\left( m \right)H}}{{{\mathbf{\tilde{\lambda }}}}^{\left( m \right)}}-{{\mathbf{s}}^{\left( m \right)H}}{{\mathbf{T}}^{H}}{{{\mathbf{\tilde{\lambda }}}}^{\left( m \right)}} \right. \\ 
     & \left. +{{{\mathbf{\tilde{\lambda }}}}^{\left( m \right)H}}{{\mathbf{\xi }}^{\left( m \right)}}-{{{\mathbf{\tilde{\lambda }}}}^{\left( m \right)H}}\mathbf{T}{{\mathbf{s}}^{\left( m \right)}} \right) \\ 
     & =\left\| \mathbf{Q}{{\mathbf{s}}^{\left( m \right)}}-\mathbf{\beta } \right\|_{2}^{2}+\frac{\eta }{2}{{\left\| {{\mathbf{\xi }}^{\left( m \right)}}-\mathbf{T}{{\mathbf{s}}^{\left( m \right)}} \right\|}_{2}^{2}} \\ 
     & +\eta \cdot real\left[ {{{\mathbf{\tilde{\lambda }}}}^{\left( m \right)H}}\left( {{\mathbf{\xi }}^{\left( m \right)}}-\mathbf{T}{{\mathbf{s}}^{\left( m \right)}} \right) \right] \\ 
     & \ge \eta \cdot real\left[ {{{\mathbf{\tilde{\lambda }}}}^{\left( m \right)H}}\left( {{{\mathbf{\tilde{\lambda }}}}^{\left( m \right)}}-{{{\mathbf{\tilde{\lambda }}}}^{\left( m-1 \right)}} \right) \right] \\ 
    \end{aligned}
       \end{equation}
       
       As a convex quadratic function, $\text{real}\left[ {{{\mathbf{\tilde{\lambda }}}}^{\left( m \right)\text{H}}}\left( {{{\mathbf{\tilde{\lambda }}}}^{\left( m \right)}}-{{{\mathbf{\tilde{\lambda }}}}^{\left( m-1 \right)}} \right) \right]$ has a lower bound. Therefore, ${{L}_{\eta }}\left( {{\mathbf{s}}^{\left( m \right)}},{{\mathbf{\xi }}^{\left( m \right)}}\text{,}{{{\mathbf{\tilde{\lambda }}}}^{\left( m \right)}} \right)$ has a lower bound, denoted as ${{b}_{L}}$.

\bibliographystyle{IEEEtran}
\bibliography{myref}

\end{document}